\documentclass[authoryear,10pt]{elsarticle}
\usepackage[applemac]{inputenc}
\usepackage[a4paper]{geometry}
\usepackage{graphicx}
\usepackage[textwidth=8em,textsize=small]{todonotes}
\usepackage{amsmath}
\usepackage{natbib}
\usepackage{hyperref}
\usepackage{placeins}
\usepackage{setspace}

\journal{Journal of Forecasting}
\begin{document}

\begin{frontmatter}
\title{Scaling-aware rating of Poisson-limited demand forecasts}
\author[add1]{Malte C. Tichy\corref{cor1}}
\ead{malte.tichy@siemensgamesa.com}

\author{Illia Babounikau}
\author{Nikolas Wolke}
\author{Stefan Ulbrich}
\author[add2]{Michael Feindt}

\address{Blue Yonder GmbH, R\"odingsmarkt 9, D-20457 Hamburg}
\address[add1]{Current address: Siemens Gamesa Renewable Energy GmbH, Beim Strohhause 17-31, D-20097 Hamburg}
\address[add2]{Current address: Deli Bringy Ltd, Al Reef St. 1, Um El Fahem, ISR-3001000}

\begin{abstract}
Forecast quality should be assessed in the context of what is possible in theory and what is reasonable to expect in practice. Often, one can identify an approximate upper bound to a probabilistic forecast's sharpness, which sets a lower, not necessarily achievable, limit to error metrics. In retail forecasting, a simple, but often unconquerable sharpness limit is given by the Poisson distribution. When evaluating forecasts using traditional metrics such as Mean Absolute Error, it is hard to judge whether a certain achieved value reflects unavoidable Poisson noise or truly indicates an over-dispersed prediction model. Moreover, every evaluation metric suffers from \emph{precision scaling}: The metric's value is mostly defined by the selling rate and by the resulting rate-dependent Poisson noise, and only secondarily by the forecast quality. Comparing two groups of forecasted products often yields ``the slow movers are performing worse than the fast movers'' or vice versa, which we call the \emph{na\"ive scaling trap}. To distill the intrinsic quality of a forecast, we stratify predictions into buckets of approximately equal predicted values and evaluate metrics separately per bucket. By comparing the achieved value per bucket to benchmarks defined by the theoretical expectation value of the metric, we obtain an intuitive visualization of forecast quality. This representation can be summarized by a single rating that makes forecast quality comparable among different products or even industries. The thereby developed \emph{scaling-aware forecast rating} is applied to forecasting models used on the M5 competition dataset as well as to real-life forecasts provided by Blue Yonder's Demand Edge for Retail solution for grocery products in Sainsbury's supermarkets in the United Kingdom. The results permit a clear interpretation and high-level understanding of model quality by non-experts.
\end{abstract}

\end{frontmatter}

\section{Introduction: The unsolved problem of forecast rating under precision scaling}
Forecasting is an application of Information Technology. Like any technology -- think of the speed of rockets or the efficiency of power plants -- forecasting performance is constrained by fundamental natural bounds \citep{Tichy_tech_2023}. This aspect is often downplayed in practice, and the unattainability of a deterministic forecast is taken as witness of forecasting problems \citep{Bower2023}. However, even assuming perfect input data quality, what a forecast can achieve in terms of precision and accuracy is not only limited by its subscriber's wishes and its creator's skills but mainly by noise: ``Forecasts characterize and reduce but generally do not eliminate uncertainty'', as distilled by \cite{probforecastingcalib}. 
A probabilistic forecast should make the strongest, yet true statement possible. In the words of \cite{GneitingKatzfuss}, ``probabilistic forecasting aims to maximize the sharpness of the predictive distributions, subject to calibration, on the basis of the available information set''. 

\subsection{Problem statement: Slow is different, baselines are needed} \label{problemstatement}
For a meaningful forecasting evaluation, it is indispensable to quantify the uncertainty that is ideally reachable under a given set of available covariates and to view the achieved forecasting performance in that context \citep{probforecastingcalib,Petropoulos2020}, for a properly chosen out-of-sample set \citep{Tashman2000}. This is, however, not what is done in practice \citep{Gartner_FA_target}. Metrics that are used in business and operations, such as Mean Absolute Error (MAE) and their normalized variants \citep{kolassa_schuetz} (errors divided by the mean observation, discussed in detail in Section \ref{metrics}) do not answer at all to which extent the forecast matches the statistical ideal of maximal sharpness under calibration \citep{probforecastingcalib,GneitingKatzfuss}, even if common pitfalls are avoided \citep{Hewamalage:2022aa,Robette2023}. Forecasters routinely ask and answer ``what MAE do I achieve?'', but, in our experience, they don't ask ``what MAE could I ideally achieve?''. 

Even skilled experts cannot develop a reliable intuition to intuitively answer the second question, due to \emph{precision scaling}: The achievable value of any metric depends on the forecasted value itself. Any metric evaluated on a group of forecasted mean values of about 1 (which retailers call ``slow-movers'') naturally averages to a different value than for a group of forecasted mean values of about 100 (the ``fast-movers'').  Metrics reflect, in the first place, the unavoidable different scale of errors for different velocities, and only secondarily the quality of the forecast. When precision scaling is ignored, one falls into what we name the \emph{na\"ive scaling trap}, the mis-interpretation of a symptom caused by precision scaling as a genuine signal of model quality. 

The problem of rating forecast quality beyond relative comparisons (on a fixed dataset, two or more forecasts are compared \citep{MAKRIDAKIS20221325,MAKRIDAKIS20221346}) therefore remains open and highly relevant. 

\subsection{Method and strategy: Split by velocity and benchmark by bucket}
Here, we elaborate on the ideas exposed in \cite{tichy2024_foresight} and solve the problem of forecast rating by a well-defined sequence of data processing and visualization techniques. These account for scaling and put the achieved values of traditional metrics into context. By explicitly answering the second question (``what MAE could I ideally achieve''), the user can then grasp the true quality of a forecast. This allows them to focus on those subsets of data that bear improvement potential instead of those that only ostensibly perform weakly. 

In a nutshell, our method consists of segregating the set of predictions and actuals into buckets that share approximately the same prediction. For each bucket, we compute the theoretical lower bound to the forecasting metric, and establish which reference values of that metric can still be considered ``excellent'', ``good'', ``OK'', ``fair'', ``insufficient'' and ``unacceptable''. The forecasting metric is then evaluated for each bucket on the actual observations and compared to the reference values. The thereby achieved ratings are aggregated to provide an overall rating (``good'', ``OK'', ``fair''...), or reference values for the achieved metric (``the overall MAE amounts to 5.4, which is `good`''). This answers the question ``what MAE could I ideally achieve?'' and sets the achieved value into that context.
 
\subsection{Literature background}
There has been much debate on the usage of the ``correct'' evaluation metric \citep{Davydenko2014,HYNDMAN2006679,Gneiting2011,Ma2018,Wheatcroft2019,Petropoulos2020,Kolassa2016,Hewamalage:2022aa}. From a practitioner's point of view, forecast quality is not a goal, but only the means to achieve the optimal business decision, which can make forecast quality unimportant \citep{on_the_selection,Robette2023,Kolassa2023,THEODOROU2025414}. Still, forecasters often need to manage stakeholders who sometimes set overly ambitious goals that are driven by gut feeling instead of quantitative benchmarks \citep{Gartner_FA_target}. 

There is a clear gap between the statistics state of the art and practice. Statisticians see forecasts as probabilistic, falsifiable statements that can be evaluated \citep{GneitingKatzfuss} using the notions of calibration and sharpness \citep{Gneiting2004,probforecastingcalib}. Such abstract notions are often avoided by practitioners. Superficially surprising or paradoxical properties of metrics, however, can often be understood via statistical concepts such as optimal point estimators \citep{Gneiting2011,Kolassa2020}. 

The effect that we focus on here, precision scaling, is due to the scaling behavior of the sharpest possible distribution in a given forecasting context. To our knowledge, precision scaling has not been investigated so far, and computing expectation values of metrics (or other statistics) under a certain distribution is not common in forecasting practice, in great contrast to, e.g., physics. Approaches to forecast evaluation that are currently used in practice, e.g.~\cite{Bower2023} proposes a ``basic heat map or listing of worst-case forecast error -- at either the item or product-family level'', are heavily infected by the na\"ive scaling trap.

For forecast evaluation, \emph{dimensionless} metrics have been proposed, which, however, do not solve the problem of precision scaling. When the popular Mean Absolute Error is divided by the mean observation \citep{kolassa_schuetz}, one obtains the dimensionless normalized MAE, sometimes dubbed Weighed Mean Absolute Percentage Error (WMAPE). Other attempts to obtain scale-free measures are MASE (mean absolute scaled error) \citep{HYNDMAN2006679}. Mean Absolute Percentage Error (MAPE) remains popular due to its superficial simplicity, despite strong criticism \citep{Kolassa2011,mape_tichy_foresight,kolassaMAPE2023}. Among other problems, MAPE is unbounded when over-forecasting, but bound by 100\% when under-forecasting, which  favors biased point forecasts. The symmetric MAPE introduced by \cite{MAKRIDAKIS1993527} resolves this asymmetry only ostensibly  \cite{GOODWIN1999405}. Other adaptations of MAPE that aim at overcoming its asymmetry  use the logarithm of the forecast-to-actual ratio \citep{Tofallis2013}. All these dimensionless metrics remain scale-dependent in the sense that they assume different values for slow and for fast velocities under a maximally sharp and calibrated forecast. 

\subsection{Forecast comparison use cases}
Different situations require users to compare forecasts:
\begin{itemize}
\item Model comparison: For a given fixed dataset, different models are compared to select the better one. 
\item Dataset comparison: For a given fixed model, subsets of data are compared, e.g., to prioritize model improvement.
\end{itemize}

Model comparison is the least problematic application of evaluation metrics: Precision scaling artefacts affect the two (or more) competing models in similar ways. When a business-relevant overall metric has improved on a given fixed dataset for a new model,  one can typically conclude that the newer model yields better performance. Nevertheless, it is helpful to be given additional information such as whether that improvement was evenly applied across all velocities. 

Dataset comparison is a routine for practitioners \citep{Bower2023} and suffers immensely from precision scaling: Given two datasets, one of them certainly contains more slowly moving items than the other. The difference in metrics then reflects this difference between the datasets, which is often erroneously interpreted as an indication for ``better`` or ``worse`` model performance. 

Scaling-aware forecast rating sheds more light on model comparison, allows dataset comparison and, to some extent, hybrid comparison across models and datasets. It is intended to allow a \emph{judgement} of forecasts by closing the gap between distribution-based statistical approaches and business applications. The reasons why the forecast on one dataset might be ``far away from Poissonian'', however, might be solvable in one case and unsurmountable in the other. We do not make statements about forecastability -- which requires a lot of situational and domain-specific knowledge -- but we provide tools to tackle domain-independent precision scaling. 

\subsection{Outline}
A sandbox model for comparing forecast accuracies  is introduced in Section \ref{sec:Evaluationmodel} along with the metrics and the M5-competition dataset \citep{MAKRIDAKIS20221346,MAKRIDAKIS20221325}. Our method is motivated and described in Section \ref{rating_by_stratification}. The technique is applied to different forecast comparison situations on the M5 dataset and on Blue Yonder's Demand Edge for Retail forecast for the UK retailer Sainsbury's in Section \ref{application}. Several possible future avenues are discussed in the Conclusions, Section \ref{sec:conclusions}.

\section{Forecasting and evaluation setup} \label{sec:Evaluationmodel}
Consider a retailer that offers many different products in many locations, for which a forecast for $n \gg 1$ different product-location-day combinations is generated. We remain with that illustrative example throughout this article. 

\subsection{A simple approximate upper sharpness bound: The Poisson distribution}  \label{poisson_sharpest}
Establishing the sharpest possible forecast under given circumstances can be challenging, but one can sometimes find approximate benchmarks. In non-personalized grocery retail, it is helpful to employ the Poisson distribution as a simple and intuitive working horse \citep{Tichy_tech_2023}: Consider the number of sold items $s$ of a product in a store on a given day, for a supermarket with $N >100$ customers per day. Our forecast predicts the number of customers $N$, and the average small probability that they buy a given product, $p < 0.05$. No customer buys two or more pieces of the same item. These $N$ independent Bernoulli-events with success probability $p$ yield the binomial probability to sell $s$ items,
\begin{equation}
P_{\text{binomial}}(s | N, p) = {N \choose s } p^s (1-p)^{N-s} ,
\end{equation}
which, for practical purposes, is indistinguishable from the Poisson distribution of rate $\mu= N \cdot p$, 
\begin{equation}
P_{\text{Poi}}(s | \mu = N \cdot p) = \frac{e^{-\mu}{\mu ^{s} }}{s!} .
\end{equation}
We assume that the forecaster performs a training with Poisson loss function and extracts the expectation value of the Poisson distribution, which is  published as a forecast for the expectation value \citep{SNYDER2012485,probforecastingcalib,https://doi.org/10.1111/j.1541-0420.2009.01191.x}.

\subsection{Corrections to the Poisson distribution} \label{corrections_poisson}
In practice, the Poisson assumption laid out above is never perfectly fulfilled. Several effects both increase and decrease the width of the ideally achievable distribution. In the first place, perfectly knowing the number of visitors $N$ and the average probability $p$  to buy an item is a strong assumption -- not all factors that influence demand can be known. Hence, this item-level buy rate $\mu=N \cdot p$ is itself randomly distributed. If that distribution is gamma, the demand distribution becomes gamma-Poisson (negative-binomial). 

The forecast could exploit personal information about customers. Instead of an average probability $p$ for the individual buy, this might be personalized to $p_1 , \dots , p_N$.  Le Cam's theorem \citep{LeCam} sets a bound to the total variation distance between the resulting Poisson-binomial and Poisson distributions. Under plausible assumptions, the potential of added forecast sharpness via personalization is severely limited: An extraordinary degree of individualization (individual customer buy probabilities need to convey a lot of information, i.e. $p_j \ll 1$ or $p_k \approx 1$) is necessary to achieve a clearly-sharper-than-Poisson distribution.

The assumption that only one item is bought per customer is also slightly broken in practice: At Sainsbury's UK, about 87.5\% of sold items are bought alone, 9.8\% are bought in pairs of two pieces, about 1.5\% in triplets of three pieces, four or more pieces are also occurring. The variance of the resulting distribution is thereby increased. In grocery retail, however, the effect of multi-buys amounts to a manageable correction.

Finite stocks censor demand values that are larger than the available stocks. This right censoring effectively decreases the width of the observed distribution. 

A priori, it is difficult to estimate the strength of these effects. For the datasets that we employed -- M5 and Sainsbury's -- we will see, {\it ex post}, that the Poisson limit is a first-order benchmark, and that the effects mentioned in this section can be considered ``second-order terms''. If a more suitable benchmark distribution is available, which may include effects such as over-dispersion, zero-inflation, and right censoring, our method can easily be adapted, {\it mutatis mutandis}.

To focus on our main contribution, the rating of forecasts across selling rates, our working assumption is that a Poisson forecast constitutes the sharpest possible prediction.

\subsection{Forecasting model}
Under the Poisson assumption, a product-location-day combination $j$ ($1 \le j \le n$) is governed by a true selling rate $t_j$, which will  remain unknown and is only under control in the numerical experiments here. The observed sales $s_j$ are samples from the Poisson distribution with rate $\mu=t_j$:
\begin{equation}
 s_j \sim P_{\text{Poi}}( s=s_j | \mu=t_j) = \frac{e^{-t_j}{t_j ^{s_j} }}{s_j!} \label{Poissondistribution}
\end{equation}
The forecast for that product-location-day $j$ is a selling rate $r_j$. The forecaster is self-confident and believes that the forecasted rates
 are perfectly known Poisson-rates, such that  each $s_j$ behaves as if it were drawn from that Poisson distribution. Our considerations are independent of \emph{how} the forecaster constructed the $r_j$: They might use a time-series approach, a machine-learning model, hybrid approaches, or any other method \citep{Petropoulos2020}.

 \subsection{Evaluation metrics}\label{metrics}
To evaluate the quality of their forecast, the forecaster relates the observed sales $\vec s=(s_1, \dots, s_n)$ to their previously predicted rates $\vec r=(r_1, \dots, r_n)$ by evaluating metrics $M(\vec r, \vec s)$. 

\subsubsection{Loss functions on point estimates}
A popular metric is Mean Absolute Error (MAE),
\begin{equation}
\text{MAE}(\vec r, \vec s) = \frac{1}{n} \sum_{j=1}^n \left| s_j - \text{median}(P_{\text{Poi}}(\mu=r_j)) \right| ,
\end{equation}
where the optimal point estimator for MAE, the median of the predicted distribution, replaces the mean $r_j$
\citep{doi:10.1080/00031305.1990.10475690,Gneiting2011}. That is, the forecaster is aware that their forecast $r_j$ is not a universal best guess suitable for every metric, but that a metric-dependent point estimate is required \citep{Kolassa2020}. MAE naturally takes larger values for larger predictions, a first attempt towards a scale-free metric is the dimensionless normalized MAE \citep{kolassa_schuetz}, often called Weighted Mean Absolute Percentage Error (WMAPE),
\begin{equation}
\text{WMAPE}(\vec r, \vec s) = \frac{\sum_{j=1}^n \left| s_j - \text{median}(P_{\text{Poi}}(\mu=r_j)) \right|  } {\sum_{j=1}^n s_j   }  = \frac{\text{MAE}(\vec r, \vec s) }{\frac 1 n \sum_{j=1}^n s_j},
\end{equation}
which turns out below to be scale-dependent as well.

The still-popular Mean Absolute Percentage Error (MAPE) comes with many flaws \citep{HYNDMAN2006679,Kolassa2011,kolassa_schuetz}. In particular, it is undefined for observations $s=0$, whose treatment heavily affects the achieved MAPE-values \citep{kolassaMAPE2023}, complemented by the particularly complicated precision scaling of MAPE \citep{mape_tichy_foresight}.

 \subsubsection{Ranked Probability Score}
MAE has an immediate business interpretation: A planner who orders $\text{median}(r_j)$ pieces will, on average, have an excess or under-stock of MAE pieces. Using the median of the distribution, however, leads to unpleasant discontinuous behavior (the median jumps from 0 to 1 at the rate $r=\log 2 \approx 0.693$). To provide a business interpretation while acknowledging the probabilistic nature of the forecast, the forecaster also evaluates the Mean Ranked Probability Score (MRPS), which is the mean over all product-location-days of the discrete Ranked Probability Score (RPS) \citep{Epstein1969,Gneiting2007,MRPS_blog}.
\begin{eqnarray}
\text{RPS}(s_j, r_j) &=& E\left( | x - s_j | \right)_{x \sim P_{\text{Poi}}(s=x | \mu=r_j) } \nonumber  \\
&& 
- \frac 1 2 E\left( |x - y | \right)_{x \sim P_{\text{Poi}}(s=x | \mu = r_j), y \sim P_{\text{Poi}}(s=y | \mu= r_j) }  , \label{MRPSdef} \\
\text{MRPS}(\vec s, \vec r) &=& \frac 1 n  \sum_j \text{RPS}(s_j, r_j) .
\end{eqnarray}
MRPS generalizes MAE from simple point estimates to distributions: For a deterministic zero-width forecast that states that $r_j$ occurs with certainty, the first term in Eq.~(\ref{MRPSdef}) becomes ${|r_j - s_j|}$ and second term vanishes \citep{Gneiting2007}. Our main working horse in this article, the Normalized Mean Ranked Probability Score (NMRPS), is then defined as
\begin{equation}
\text{NMRPS}(\vec s, \vec r) = \frac {\text{MRPS}(\vec s, \vec r)} { \frac 1 n  \sum_j s_j} = \frac{\sum_j \text{RPS}(s_j, r_j)} {\sum_j s_j} .
\end{equation}

Precision scaling is ubiquitous, it affects (N)MRPS and all other metrics in a similar way as MAE/WMAPE.

\subsection{Baseline model and na\"ive scaling trap} \label{benchmark_model}
As test-bed of our method, we use the dataset for the validation period, 2016-04-25 to 2016-05-22, of the M5-competition  \citep{MAKRIDAKIS20221325,MAKRIDAKIS20221346}. To obtain a maximally sharp and calibrated placeholder for a perfect baseline forecast, we construct predictions $r^{\text{baseline}}_j$ via in-sample Expectation Maximization (EM) of the Poisson likelihood, applied on the entire set of sales $\vec s$ (see Supplemental Material). In short, the Poisson-rates $r_j$ that are generated by this procedure could plausibly have have produced the outcomes $s_j$. 
 
 A forecaster is interested in the overall performance of their model (``is that a good model?''), and in systematic patterns in the performance (``are we doing better in Hobbies or in Household?''), and therefore evaluates the forecast not only globally, but also segregated by department (or by other properties known at the moment of the forecast). 
 
In Table~\ref{KPI_benchmark}, for the total, and for each of three selected departments (``Household 1'', ``Hobbies 2'', ``Foods 3''), all achieved metrics are finite for the baseline forecast. Remember that the forecast was fabricated to be ideally Poissonian for the entire assortment -- by construction, it has the same quality in every subset of data. The result of a correct analysis should produce the result ``there is no significant difference in forecast quality between the departments''. Hence, the following intuitive conclusions are erroneous: ``We need to improve Foods 3! Fast-movers are performing worse than slow-movers!'' -- even though it's corroborated by MAE and MRPS, ``We must improve Hobbies 2! Slow-movers are performing worse than fast-movers!'' is backed by WMAPE and NMRPS. The striking inconsistency of these conclusions shows that the judgement of forecast quality across departments is not straightforward. 

A direct interpretation of the forecast metrics across any segregation of the data  \citep{Bower2023} is deeply problematic and is doomed to misalign efforts for forecast improvement. Since the ``effect'' that one sees is infected by precision scaling, we propose to designate such distortion as the \emph{na\"ive scaling trap}.

\begin{table}
\begin{tabular}{r| r | r r r r r r r}
		 & All & Foods 3 & Hobbies 2 & Household 1\\ \hline
Number of predictions $n$ &853'720&230'440&41'720&148'960 \\
Total sales $\sum_j s_j$&1'231'764&564'926&13'302&222'327 \\
Total prediction $\sum_j r_j$&1'236'224&566'248&13'469&223'244 \\
Mean sales $\sum_j s_j / n$  &1.44 & 2.45& 0.32& 1.49 \\
Bias factor $\sum_j r_j / \sum_j s_j$ &1.0036&1.0023&1.0126&1.0041 \\
MAE&0.653&0.901&0.244&0.740\\
WMAPE&0.453&0.368&0.766&0.496 \\
MRPS&0.461&0.633&0.182&0.520 \\
NMRPS &0.319&0.258&0.570&0.348 \\
\end{tabular}
\caption{\label{KPI_benchmark}Metric values for baseline model. The level of confidence that is expressed by these metrics values without context depends on the interpretation of the forecast user.}
\end{table}

\FloatBarrier

\section{Method: Rating metrics in a scaling-aware way}  \label{rating_by_stratification}

\subsection{Revealing and avoiding the na\"ive scaling trap}
The reason why the metrics in Table~\ref{KPI_benchmark} amount to different values in the different departments lies in the scaling of the variance of the Poisson distribution with the rate $\mu$: Each individual prediction $r_j$ comes with a different expectation of how $s_j$ should be distributed (Eq.~(\ref{Poissondistribution})), and which metric value $M(r_j, s_j)$ should be achieved on average. The widespread assumption that the same metric value is achievable for each category is therefore flawed. 

It is useful to compute the expected value of the metric, given the predictions, under the assumed sharpest possible forecast. These ideal values constitute a baseline comparison for the actually achieved metrics. A marginal discrepancy would confirm that the forecast is almost as good as possible. A large difference motivates an investigation whether it is possible to close that gap. 

If all predictions are aggregated globally, the resulting expectation value of the metric will be mostly borne by fast-movers. We therefore segregate by selling rate: We group the pairs of predictions $r_j$ and observations $s_j$ into buckets, the bucket of a pair is chosen according to the prediction value $r_j$. Choosing the outcome $s_j$ as identifier would induce the hindsight selection bias \citep{hindsight_bias_blog,LerchForecastersDilemma,preventingforecastersdilemma,tichy2024forecastersdilemma}. Logarithmically spaced buckets avoid that one bucket contains the majority of all predictions, i.e.,~we group by
\begin{equation}
R_j =  \frac 1 { n_{\text{bins}} } \text{round}\left( n_{\text{bins}} \text{log}_{10}(r_j) \right) , \label{roundedlogarithm}
\end{equation}
where we round to integers such that there are $n_{\text{bins}}$ bins between two powers of 10. For example, for $n_{\text{bins}}=4$, the $R_j$ that can take the values $0, 0.25, 0.5, 0.75, 1, 1.25$ etc. Buckets are referred to by their common rounded logarithmic prediction value $R$. The size and number of the buckets should then ensure a good balance between precision (many small, homogenous buckets) and sufficient statistics per bucket (few large more inhomogeneous buckets); in practice, $n_{\text{bins}} \approx 5$ is a reasonable choice.

The predictions $r_j$ that belong to the same bucket, indexed by $R$, come with similar expectation values, for any metric. Ideally, when the prediction is truly Poissonian, we expect the achieved average metric in a bucket to match the metric's computed expectation value. 

\begin{figure}
\centering
\includegraphics[width=0.5\textwidth]{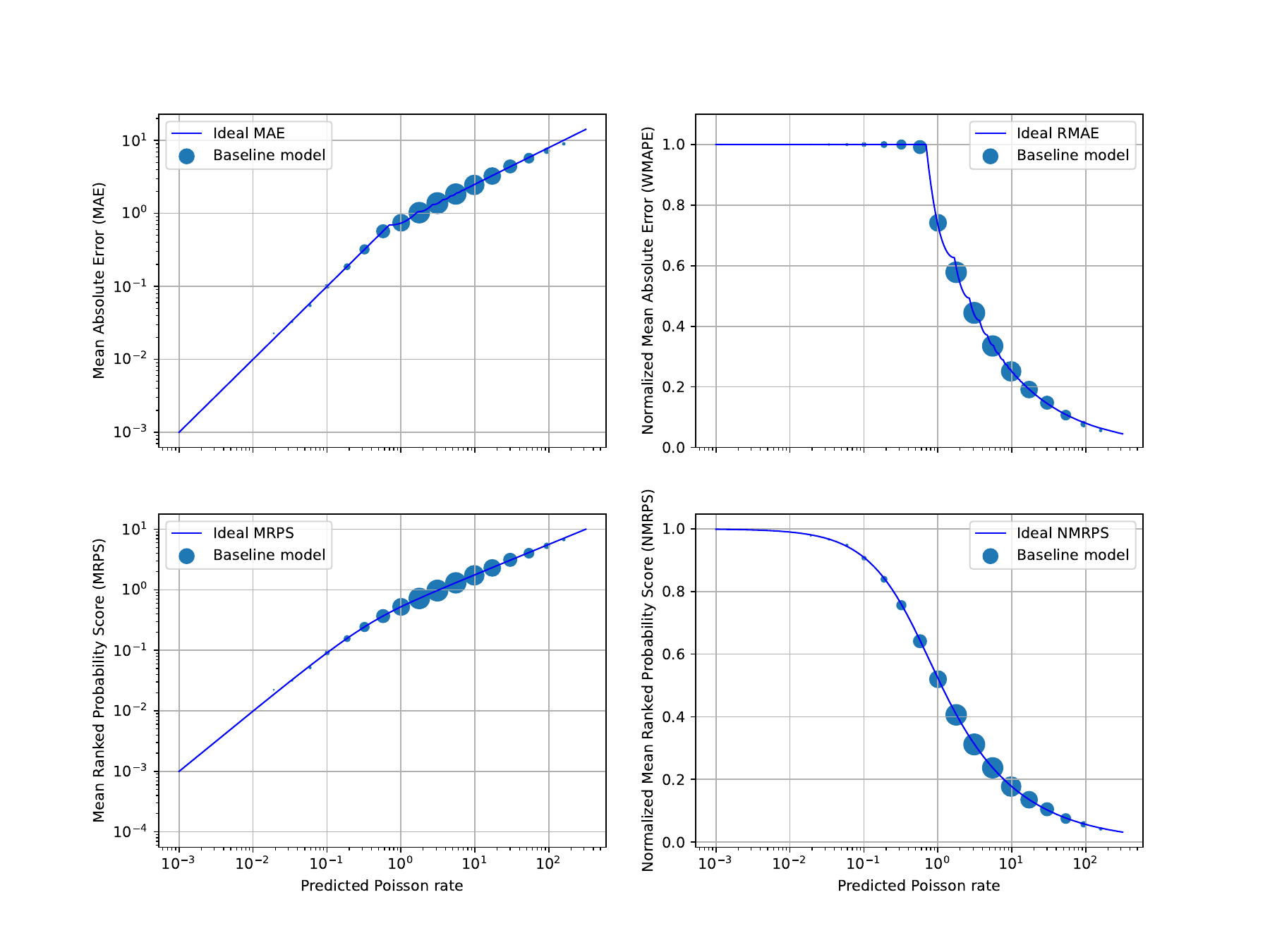}
\caption{ \label{figscatter_benchmark} Ideal Poisson NMRPS (solid lines) and achieved NMRPS for baseline model (circles). The size of the circles reflects the square root of the number of sales grouped in one bucket. }
\end{figure}

\begin{figure}
\centering
\includegraphics[width=0.87\textwidth]{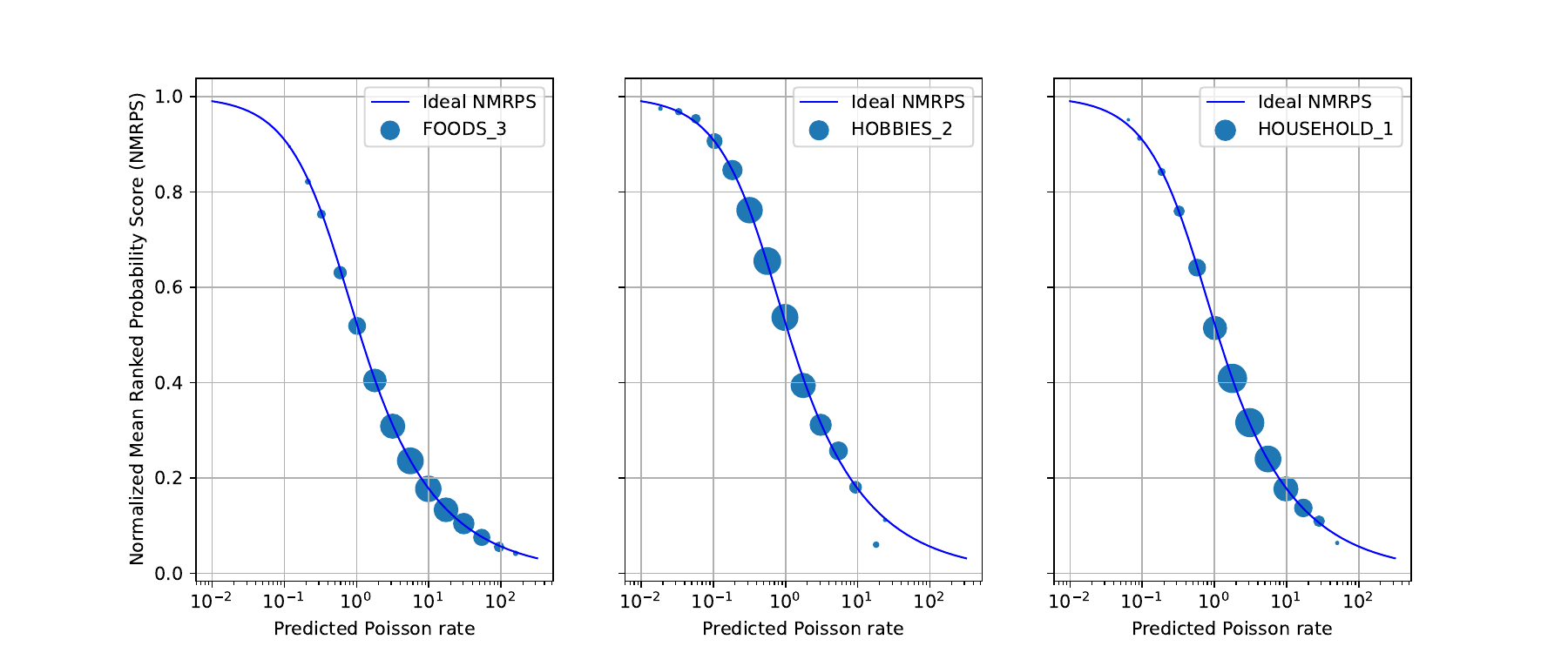}
\caption{ \label{figscatter_per_department} Poisson-achievable NMRPS (solid lines) and achieved NMRPS (circles) for three selected departments, for the baseline model. While the overall NMRPS per department differs substantially (see Table~\ref{KPI_benchmark}), the differentiation by prediction bucket shows that the performance is equally good.}
\end{figure}

Figure \ref{figscatter_benchmark} illustrates the result of this procedure for the baseline model as circles and $n_{\text{bins}}=4$, and superimposes the expectation value of the NMRPS under the Poisson distribution as solid lines. The positions of the circles match the  lines, confirming the Poisson performance of our baseline model. Figure \ref{figscatter_per_department} differentiates further by retail department: The differences in the metric values exhibited in Table \ref{KPI_benchmark} are driven by the differences in the distribution of rates among the departments, not by a different performance. For a given rate bucket $R$, the performance is the same for all departments, but the different buckets have different populations, which leads to a different overall metric value. Skepticism that calls into question the equal model performance in the different departments, which might have been induced by the different metric values in Table \ref{KPI_benchmark}, can now be rejected.

The intricate shapes that the NMRPS exhibits as a function of the predicted rate in Figure \ref{figscatter_benchmark} makes comparisons of traditional metrics between strata of data \citep{Bower2023} deeply problematic. The ``best case'' value of a metric depends strongly on the prediction itself. Faster-selling departments naturally achieve higher values of MAE and MRPS, and lower values of WMAPE and NMRPS, as exemplified by Figure \ref{figscatter_per_department}. Therefore, statements like ``we achieve 38\% WMAPE on Foods 3, but 78\% WMAPE on Hobbies 2, we need to improve Hobbies 2'' are infected by the na\"ive scaling trap and need to be dismissed.

%\FloatBarrier

\subsection{Imperfect reference forecasts} \label{scaling_imperfect_reference}
The setting so far was artificial: The baseline model is ideally Poissonian and unsurprisingly matches the theoretical expectation. We merely confirmed that the Expectation Maximization algorithm works. Figure \ref{figscatter_per_department} explains why model performance across departments is only ostensibly different. Characterizing the ideal case is, however, not sufficient for judging forecasts in practice: One needs to quantify and rate how close the model performance is to ideal. For this purpose, we define below use-case dependent benchmark forecasts that set the standard for ``Excellent'', ``Good'', ``OK'', ``Fair'', "Insufficient" and ``Unacceptable'' performance, both regarding the level of bias and the level of noise.

Perfect performance is uniquely defined, there is one way to be ``right'': Given a Poisson-noise-limited situation, the best possible value of any metric is the value achieved for outcomes drawn from the Poisson distribution. For the bias (the quotient mean prediction / mean observation), the ideal value is one. For other metrics, analytic formulae are often available (see \cite{Wei2014} for WMAPE/MAE and (N)MRPS); otherwise, one can recur to Monte Carlo sampling. 

There are, however, many ways to be wrong: A forecast can be biased, it can be affected by noise, suffer from both, or by more complicated artefacts. No general ``worst benchmark model'' can be defined for unbounded metrics: Given a forecast, one can construct arbitrarily incompatible (``bad'') observations; given a set of observations, one can construct arbitrarily terrible forecasts. 

\begin{figure}
\centering
\includegraphics[width=.6\textwidth]{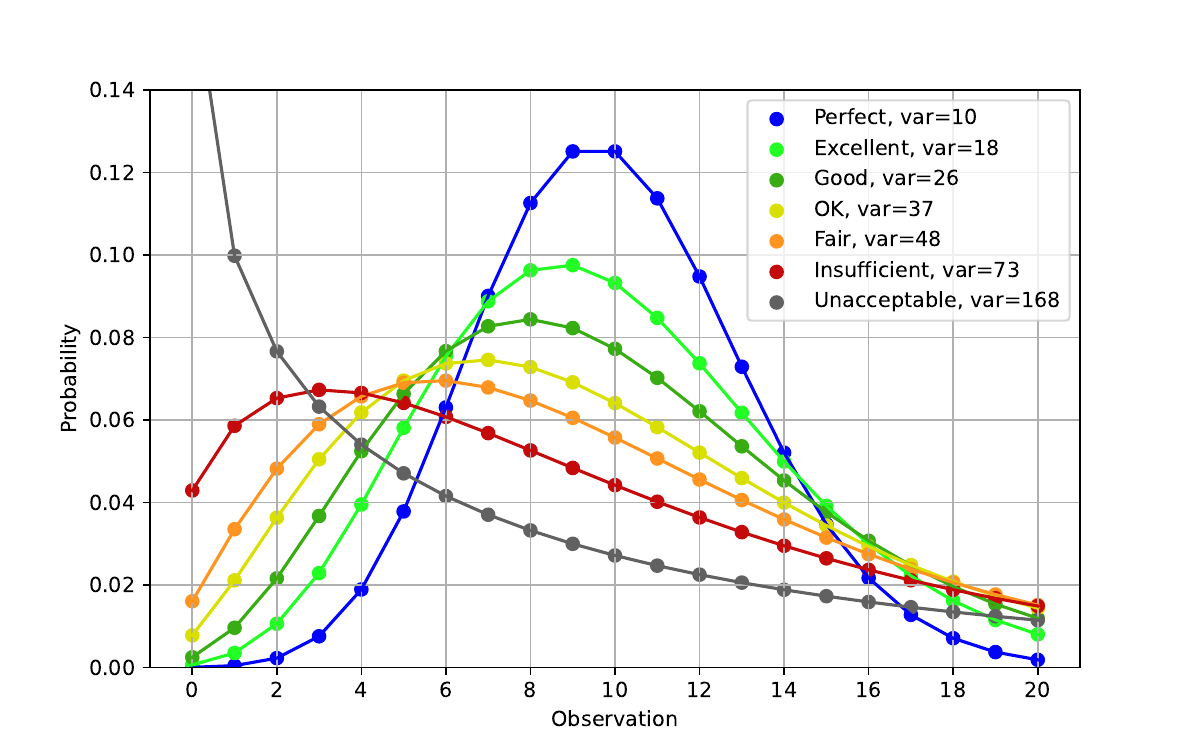}
\caption{ \label{benchmarks} Over-dispersed distributions that define the reference forecasts ``perfect'' (Poissonian, blue), ``excellent'' (light green), ``good'' (dark green), etc., for a  predicted rate  $\mu=10$. When the forecaster predicts a Poisson-distributed target, we rate their performance with respect to a metric by reference to the metric values achieved by these ``excellent'', ``good'', etc. negative binomial distributions.}
\end{figure}

While there is no unambiguous way to characterize imperfect forecasts, the following parametrization provides meaningful references: The forecaster continues to produce a Poisson-forecast $\mu=r_j$, but the observations $s_j$ are drawn from negative-binomial distributions fulfilling
\begin{equation}
\text{variance} = \mu + f \mu^\gamma , \label{variance_general_scaling}
\end{equation}
where the over-dispersion factor $f \ge 0 $ reflects the quality of the forecast ($f=0$ is the Poisson baseline). As we argue below in Section \ref{gammascaling}, it is plausible to set  $\gamma=1.5$. In other words, the forecaster underestimates their forecast's uncertainty; they observe an unexpectedly over-dispersed distribution of outcomes.  We now assign different scores to different degrees of over-dispersion $f$ in Figure \ref{benchmarks}. The factors $f$, the variances at prediction $r_j=10$ and the maximal overall bias factors (mean prediction / mean observation) that are rated with a certain verdict are summarized in Table \ref{scoreperquality}. 

\begin{table}
{\small
\vspace{0.35cm}
\begin{tabular}{r|lllllll}
Rating & Perfect & Excellent & Good & OK & Fair & Insufficient & Unacceptable \\ \hline
Over-dispersion factor $f$ & 0 & 0.25 & 0.50 & 0.85 & 1.2 & 2 & 4 \\
Variance at $r_j=10$ & 10 &  18 &  26&  37 & 48 & 73 & 136 \\  \hline
Bias factor & 1.0 & 1.015  & 1.03 &  1.07 & 1.2  & 2 & 4  \\
\end{tabular}
\vspace{0.35cm}
}
\caption{\label{scoreperquality} Over-dispersion, variance and bias values that are considered ``Perfect'', ``Excellent'', etc., in a rating scheme that is adapted to the M5 competition. The table is read as follows: For a prediction of $r_j=10$, an observed variance between 26 and below 37 is considered ``Good''; a variance above 136 is rated ``Unacceptable''; an overall bias (mean prediction / mean observation) between 1.2 and below 2 is judged ``Fair''.}
\end{table}

The 12 parameters (6 over-dispersion and 6 bias factors) fully define our rating, they were chosen such that a na\"ive collective global forecast for the M5 competition data is rated ``Insufficient'' by most metrics. We recommend create a rating by defining the ``Insufficient'' rating in a way that a na\"ive base model would be within this rating. The ratings for bias and for variance are independent, that is, a forecast can enjoy Poisson-like variance, but be quite biased, or vice versa. 

\subsection{Scaling of reference metric values} \label{gammascaling}
For a given metric $M$, we need to compute reference values $M_{\text{quality}}^{~(R)}$ for each quality (``Excellent'', ``Good'' etc.), and for each bucket of predictions $\vec r^{~(R)}$. The achieved value $M_{\text{actual}}^{~(R)}=M(\vec r^{~(R)}, \vec s^{~(R)})$ can then be set into that context to rate each individual bucket. We have done this exercise for one value of $R$ in Table \ref{scoreperquality}. 

\begin{figure}
\centering
\includegraphics[width=.9\textwidth]{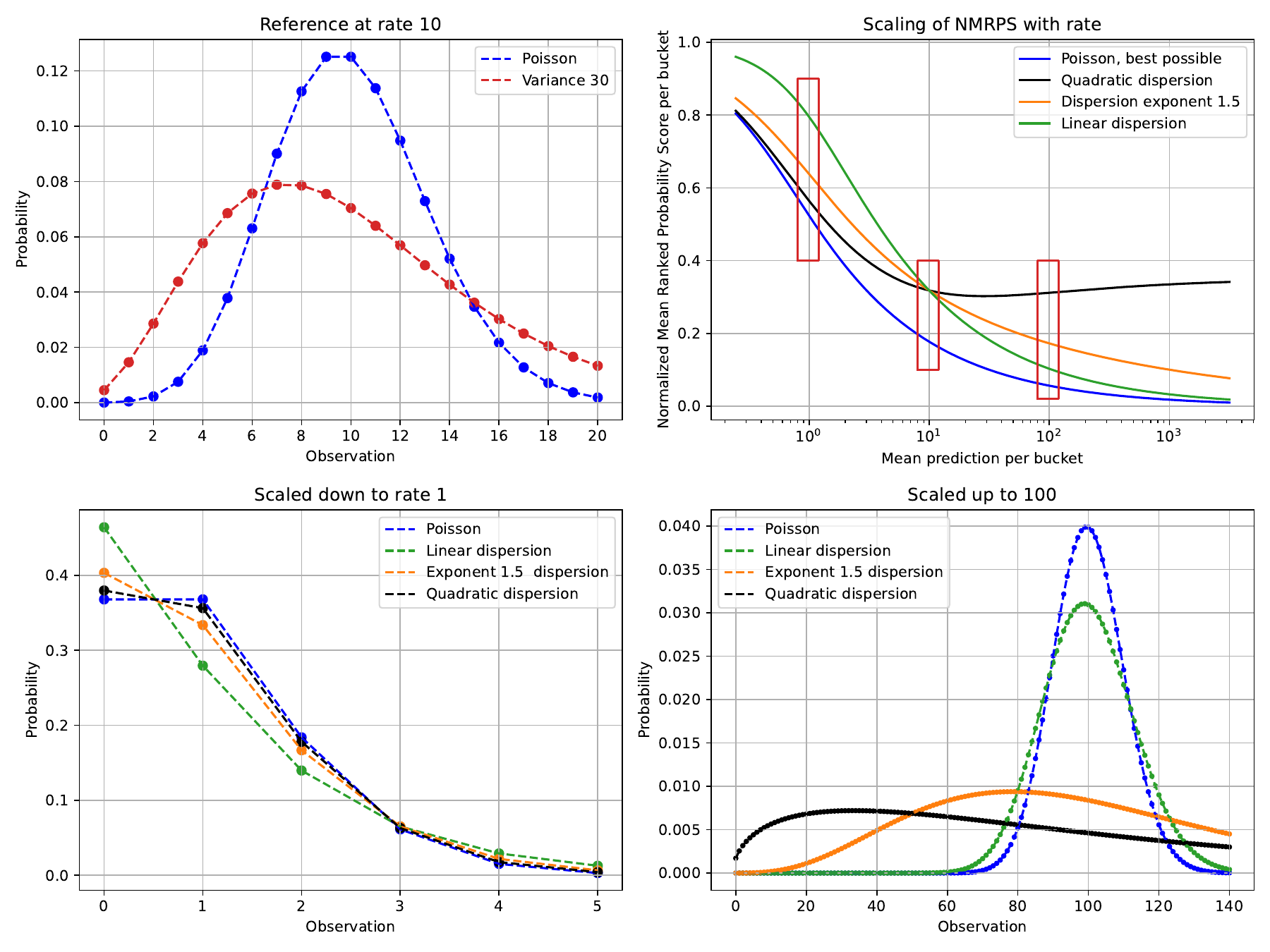}
\caption{ \label{ratingscaling} Upper left: Poisson (blue) and negative binomial (red, $\text{variance}=30$) distributions for rate $\mu=10$. Upper right: expected NMRPS as a function of rate, when the variance equals 30 at prediction 10 (corresponding to ``good'' as defined in Figure \ref{benchmarks}), for over-dispersion exponents $\gamma=1$ (green), $\gamma=1.5$ (orange), $\gamma=2$ (black). The blue line delimits the ideally achievable (Poissonian) baseline value under over-dispersion 0. Lower left: rate-10-distribution scaled down to rate 1, using linear (green), quadratic (black) and exponent-1.5 (orange) dispersion. Lower right: rate-10-distribution scaled up to rate 100, using linear (green), quadratic (black) and exponent-1.5 (orange) dispersion.}
\end{figure}

When moving to lower or higher predictions, the question arises how the variance should scale when the quality remains the same: Having established that for rate of 10, a variance of around 30 is judged as ``good'', which variance is ``good'' when the prediction is 1, 100, or 10'000? The variance at the other rates is determined by the exponent $\gamma$, illustrated in Figure \ref{ratingscaling}. Setting $\gamma=2$ corresponds to quadratic over-dispersion (a Negative Binomial-2-process), and $\gamma=1$ to linear over-dispersion (a Negative Binomial-1-process) \citep{Hilbe2012,camerontrivedibook}. The upper right panel of Figure \ref{ratingscaling} shows the NMRPS as a function of the predicted rate, for different exponents $\gamma=1, 1.5, 2$. Quadratic $\gamma=2$ (black line) is unrealistic: It is too benevolent for large rates and too strict for small rates. $\gamma=1$ (green line) exhibits the opposite behavior: It judges large rates strictly and small rates too benevolently. The intermediate value $\gamma=1.5$ (orange) provides a reasonable compromise, corroborated by this argument: When summing negative binomial-distributed random numbers, $\gamma=1$ corresponds to uncorrelated noise, $\gamma=2$ to perfectly correlated noise, and $\gamma=1.5$ to partial correlation. This is realistic in the retail setting: The number of visitors of a supermarket has some uncertainty that leads to correlated, product-independent noise on the demand, but also each individual product suffers from uncorrelated, product-dependent noise. 

\begin{figure}
\centering
\includegraphics[width=0.5\textwidth]{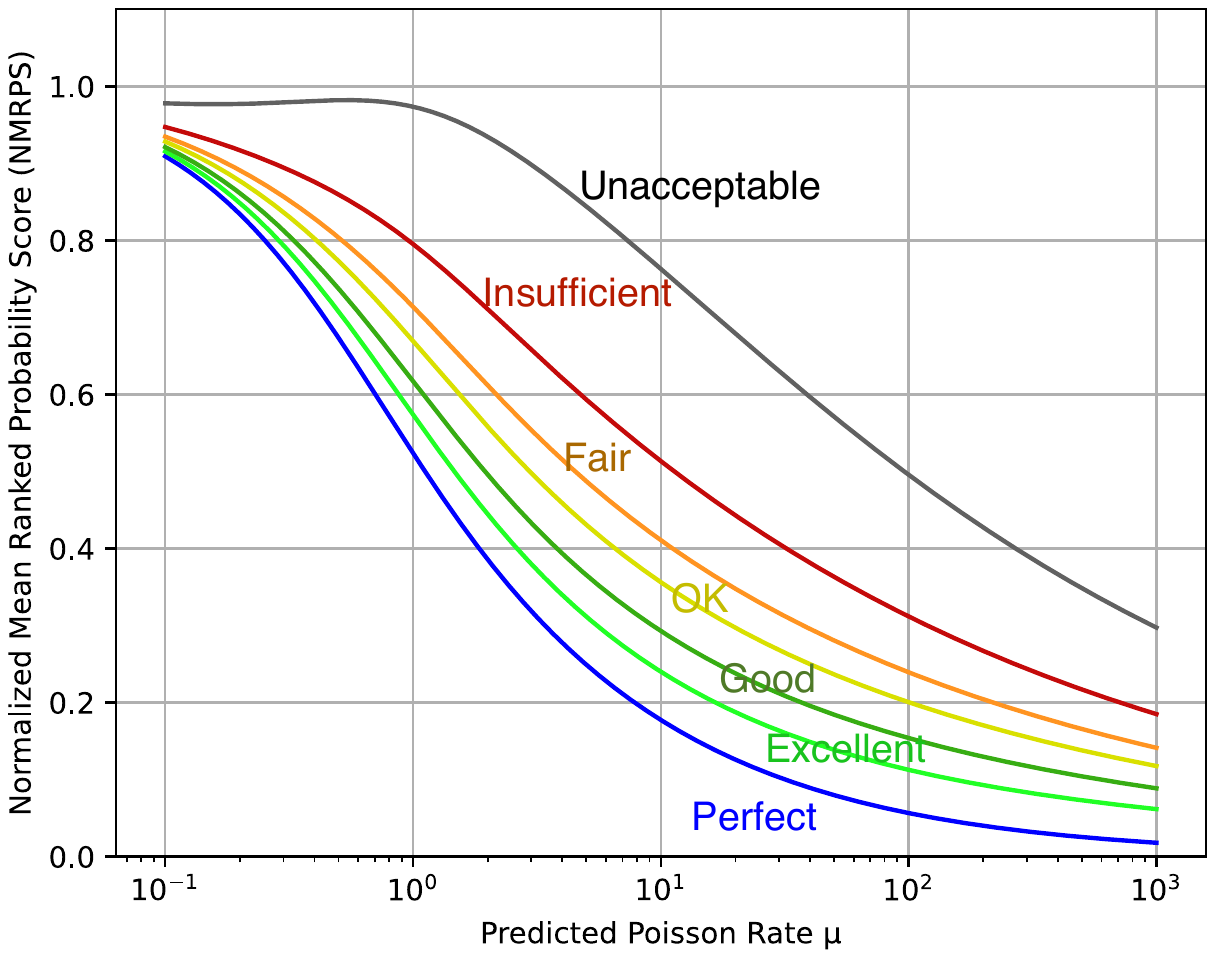}
\caption{ \label{fig9} Reference values of NMRPS as a function of predicted rate, for the different scaling-aware quality ratings. We use the variance values defined for $\mu=10$ for the different qualities in Figure \ref{benchmarks}, and apply the scaling rule Eq.~(\ref{variance_general_scaling}) with the exponent $\gamma=1.5$ to infer the variance at rates $\mu\neq 10$. This figure answers the following question: ``What value of $M$ is excellent/good/OK... for a prediction $\mu$?''}
\end{figure}

The resulting rate-dependent reference values for the exponent $\gamma=1.5$ are shown in Figure \ref{fig9}.  For a given prediction bucket, one reads off what the achieved metric means in terms of model quality (``Excellent'', ``Good'', etc.) by locating the bucket in the respective plots. By assigning ``perfect'' the numerical score 100\% and ``unacceptable'' the score 0\%, and interpolating between reference lines, each bucket $R$ gets assigned a score $S^{(R)}$. We will see below that empirical real-life forecasts match the behavior described by $\gamma=1.5$ well. 

A given metric value, e.g., NMRPS=0.5, can be considered ``perfect'' for a predicted rate of 2, or ``unacceptable'' for a predicted rate of 100. By applying the method to different sets of data from different industries, we have experienced that ``large'' buckets (containing many, more than 100, predictions and observations) that are found outside the ``insufficient'' region are indeed rare and point the user to severe model or data quality issues. 
 An unexpected metric value that a small bucket assumes, on the other hand, is not necessarily significant. 

\FloatBarrier

\subsection{Summarizing bucket-wise to overall ratings} \label{summarizing}
It is desirable to summarize the information contained in a set of buckets by a single number. There are two possibilities to perform this summary, discussed in the Supplemental Material: One can aggregate the ratings per bucket via taking the mean of the bucket-wise scores $S^{(R)}$ (weighted by the number of observations per bucket), which yields an overall score $S_{\text{overall}}$ between 0\% (all buckets are ``unacceptable'') and 100\% (all buckets are ``perfect''). Alternatively, one can judge the overall aggregated metric, e.g.,~the overall achieved WMAPE, by setting it into the context of the metric that would have been achieved if the model quality were ``excellent'' to ``good''. In other words, one can compute the expectation value of the metric under reference models of imperfect quality, answering the important question ``what values of the metric could I possibly achieve?'', which we had put forward above in Section \ref{problemstatement}.

\subsection{Ostensibly better-than-Poisson measurements} \label{ostensibly}
One will sometimes find buckets $R$ that achieve a ``better-than-Poisson'' metric, $M^{~(R)} < M^{~(R)}_{\text{perfect}}$. This unexpected behavior can be due to different causes:
\begin{itemize}
\item Insufficient statistics: The significance of such ``better-than-Poisson''-measurement is jeopardized by a too small number of prediction-observation-pairs in a given bucket. 
\item Sub-Poissonian process: The Poisson assumption might be too pessimistic. 
\item Finite capacity: If stocks are finite, not all demand peaks are reflected by corresponding sales values, but censored. Fewer events with $s_j \gg r_j$ will then be observed than were predicted, and, consequently, error metrics achieve lower values. 
\item Systematic over- or under-prediction: When a bucket is systematically over- or under-predicted ($r_{\text{total}}^{~(R)} > s_{\text{total}}^{~(R)} $ or $r_{\text{total}}^{~(R)} < s_{\text{total}}^{~(R)}$) the resulting position in the rate vs. metric plot will be shifted to the left or right, possibly ending at a ``better than Poisson'' position. A good noise rating for a bucket should only be taken as a sign that the forecast is good if the bucket is also unbiased.
\item Overfitting: When rating a model in-sample, a sub-Poissonian performance can witness overfitting. E.g., trivially setting $r_j=s_j$ gives an implausible WMAPE of 0.
\end{itemize}
Depending on which of these non-exclusive causes are realistic in a given application, one can decide to rate buckets for which $M^{~(R)} < M^{~(R)}_{\text{perfect}}$ with some critical quality to ensure that these buckets do not push the overall rating to good values, obfuscating some problems. 
The method of rating the overall aggregated metric is especially prone to cancellation between smaller-than-ideal and larger-than-ideal values, which would result in an overall ``excellent''-looking metric. Just like negative and positive bias in slow- and fast-movers may cancel to yield an ostensibly good global bias, ``better-than-Poisson'' and ``worse-than-Poisson'' buckets may result in an ostensibly good overall noise. Therefore, we prefer to aggregate the bucket-wise-scores as a primary tool of investigation.

\FloatBarrier

\section{Applications} \label{application}

\subsection{M5-competition models}
We apply our scaling-aware rating to several models for the validation period of the M5 competition \citep{MAKRIDAKIS20221346,MAKRIDAKIS20221325,kaggle}.
\begin{itemize}
\item {\bf Baseline.} The ``perfect'' baseline model described in Section \ref{benchmark_model} (details in the Supplemental Material) is calibrated and maximally (Poisson) sharp. 
\item {\bf Na\"ive-1-day-ahead.} A mediocre model uses a one-day-ahead heuristic that takes yesterday's observed sales value of every product-location combination as prediction for today's sales, effectively a one-day-horizon.
\item {\bf Simple-28-day.} This simple model \citep{simple28days} uses the last 28 days and averages the sales per weekday for each product. It predicts on horizon one (first day of the validation period) to horizon 28 (last day of the validation period).
\item {\bf Global forecast.} An undifferentiated collective forecast produces the overall average of the Simple-28-day-model for every SKU and every day. 
\item {\bf LightGBM.} The LightGBM model \citep{lightgbm} implements a tree-based learning algorithm within the gradient boosting framework \citep{ke2017lightgbm}. It was trained with a Poisson objective and RMSE as a metric, and uses some additionally engineered features such as lagged sales. It predicts on horizon one to horizon 28, and scored an above-average performance on the public leaderboard. 
\end{itemize}

To rate the models, we follow the procedure described above in Section \ref{rating_by_stratification}: We bucket predictions by their value, and plot the bias and the NMRPS for each bucket in Figure \ref{m5application}. We clip small predicted rates at 0.01 and we clip the bias factor at 10 and 1/10  to fit into the visual representation.

\subsubsection{Model comparison: Rate-stratified representation}
\begin{figure}
\centering
\includegraphics[width=0.87\textwidth]{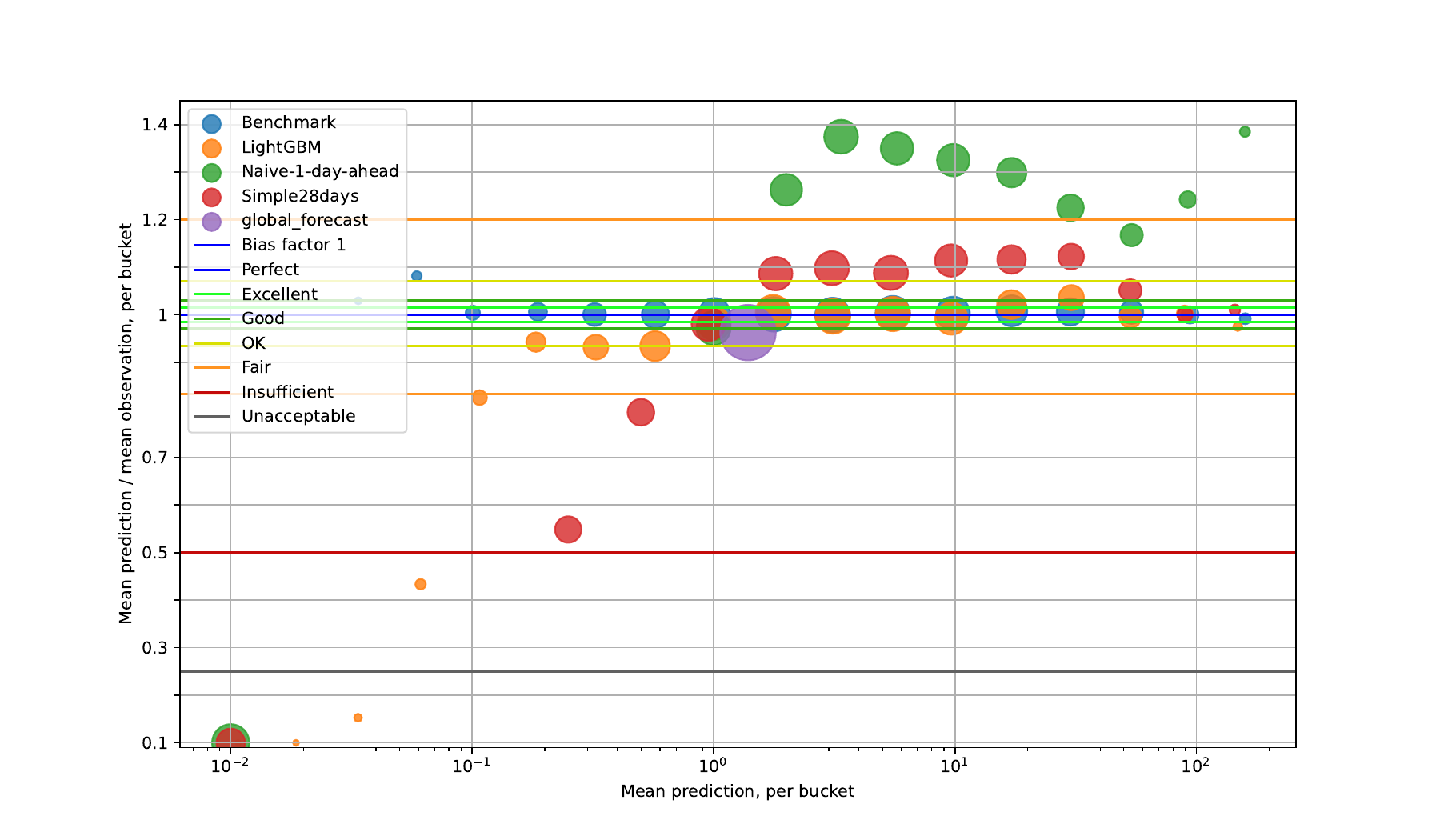}
\includegraphics[width=0.87\textwidth]{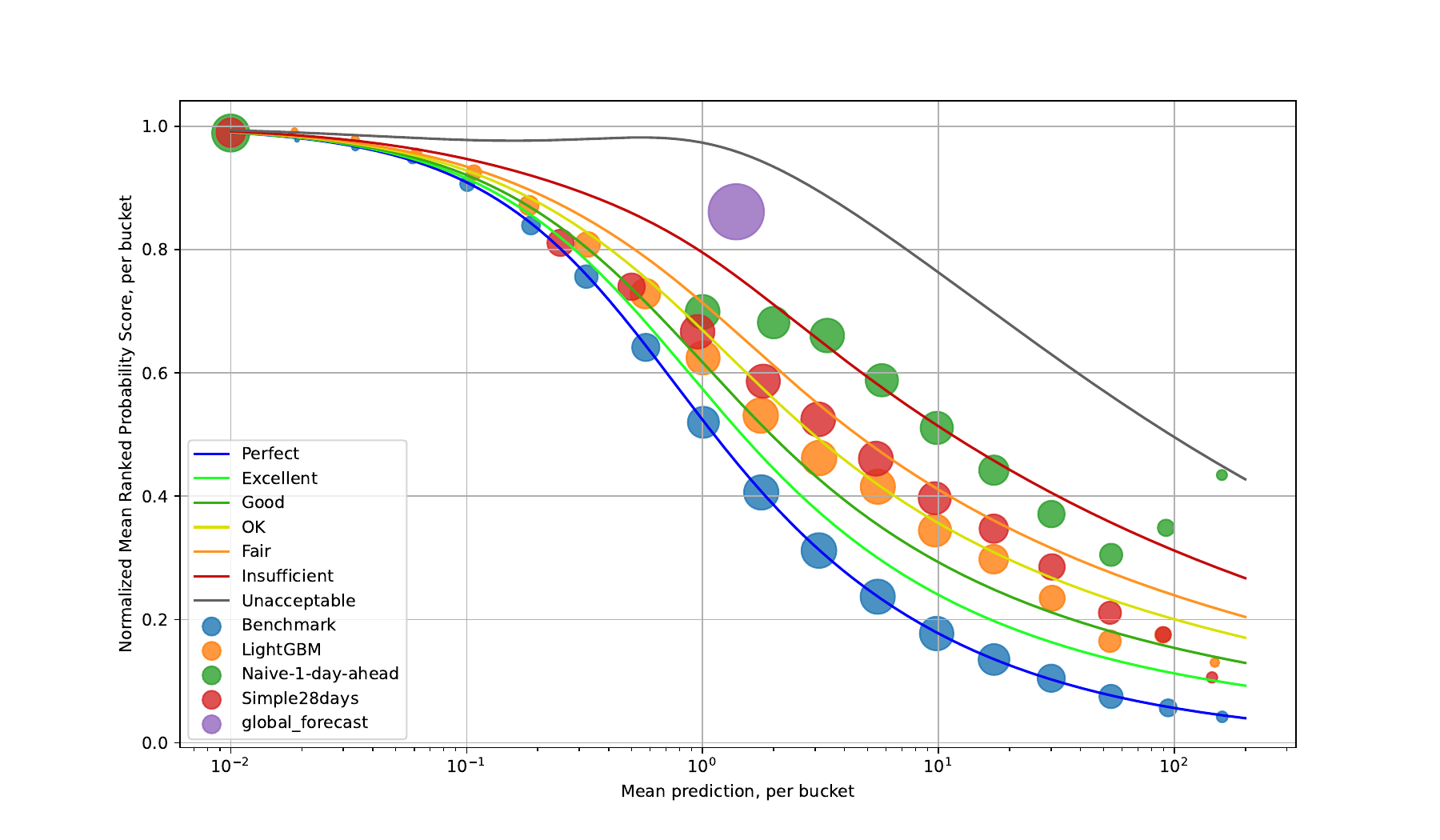}
\caption{ \label{m5application} Prediction-bucketed representation of bias (upper panel) and noise (NMRPS, lower panel) for five exemplary models. The size of the circles reflects the total number of sales contained in the respective buckets.}
\end{figure}

In the bias scatter plot in the upper panel of Figure \ref{m5application}, the baseline (blue) is almost perfect by construction, and LightGBM (orange)  exhibits a good level of bias for moderate predicted rates of 1 and above. For smaller rates ($r<0.1$), LightGBM-predictions  are typically under-predictions, witnessing overfitting. The Na\"ive-1-day-ahead prediction (green) contains many 0-predictions (which were clipped to 0.01), which are often under-forecasts: When a product is not sold on a given day, the na\"ive-1-day-ahead-prediction for the next day is zero, but the expected value of its sales for the next day is larger than zero. This regression-to-the-mean also manifests in forecasts that follow days with abnormally many sales, these are then typically over-predictions. The green circles consequently all lie above the ideal line for predictions larger than 1 (the overall average daily selling rate is about 1.4). The Simple-28-day-model is not as unbiased as LightGBM, but it is still less biased than the na\"ive-1-day-ahead model. The global model produces 1.39 for every SKU, a slight under-forecasting, since the overall mean sales are about 1.44.

The stratified noise plot in the lower panel of Figure \ref{m5application} clearly separates the five models: The baseline fits the Poisson-ideal perfect line (reproducing Figure \ref{figscatter_benchmark}). For every prediction bucket, LightGBM is more noisy, rated in the ``good'' range. The Simple-28-day model is, again, slightly worse, populating the ``OK'' area. The scattered circles follow the parametrized corridors with $\gamma=1.5$ remarkably well. 
The na\"ive-1-day-ahead model follows with some distance to the others on the boundary to ``insufficient''. The largest two prediction  buckets of that model leave the bulk -- the regression to the mean is especially pronounced for the largest predictions, which correspond to the largest observations on the day before. Not surprisingly, the collective global model exhibits a large degree of noise, close to ``unacceptable''. 

Even though NMRPS is superior to WMAPE in its handling of slowly moving predictions, it nevertheless does not allow us to differentiate between models in the ultra-slow-regime well, where all the lines converge. It remains imperative to always consider bias \emph{and} noise together.

Table \ref{ratingM5} summarizes Figure \ref{m5application} via the scaling-aware ratings, computed following Subsection \ref{summarizing} (see also the Supplemental Material). With Figure \ref{m5application} and the resulting ratings at hand, one gains the confidence that the LightGBM-model is consistently better than Simple-28-days, across all prediction buckets and in a stable fashion. The rating of bias per bucket is more stringent than the overall bias alone: The Na\"ive-1-day-ahead method is, by construction, overall unbiased (it uses the sales of the last day before and of the first 27 days of the validation period), but only the bucket with $R \approx 1$ is unbiased, resulting in a ``Fair'' overall bias performance. The ``insufficient'' rating of the global collective model reflects its lack of any individualisation among SKUs and days.

\begin{table}
{\footnotesize
\begin{tabular}{l|ll|ll|l}
Model &  $S_{\text{overall}}$(NMRPS)  &   Quality & $S_{\text{overall}}$(Bias) & Quality   & Bias factor \\ \hline
Baseline &  99.9 \%& Perfect & 98.2  \%& Perfect & 0.9937 \\
LightGBM & 66.4 \%&  Good & 87.6  \% & Excellent & 0.9904 \\
Simple-28-day &  57.4 \%& OK  &53.1 \% & OK & 0.9626 \\
Global & 16.4 \% & Insufficient & 71.3 \% & Good &   0.9626  \\
Na\"ive-1-day-ahead & 41.0\%& Fair  &  36.9 \% & Fair &1.0001 \\
\end{tabular}
}
\caption{\label{ratingM5} Scaling-aware ratings for the different models employed on the M5-competition dataset. The ratings summarize the predicted-bucketed representation of Figure \ref{m5application}.}
\end{table}

\FloatBarrier

\subsubsection{Dataset comparison: Department performance}  \label{sec:dataset_comparison}
Scaling-aware model rating adds more details to model comparison, but often reproduces the verdict of scaling-unaware metrics. For dataset comparison, the added value of scaling-aware rating is more evident: Here, the conclusion $M_A < M_B \Rightarrow $ ``A is better than B'' must not be drawn, because it is almost always affected by precision scaling. 

\begin{figure}
\centering
\includegraphics[width=0.87\textwidth]{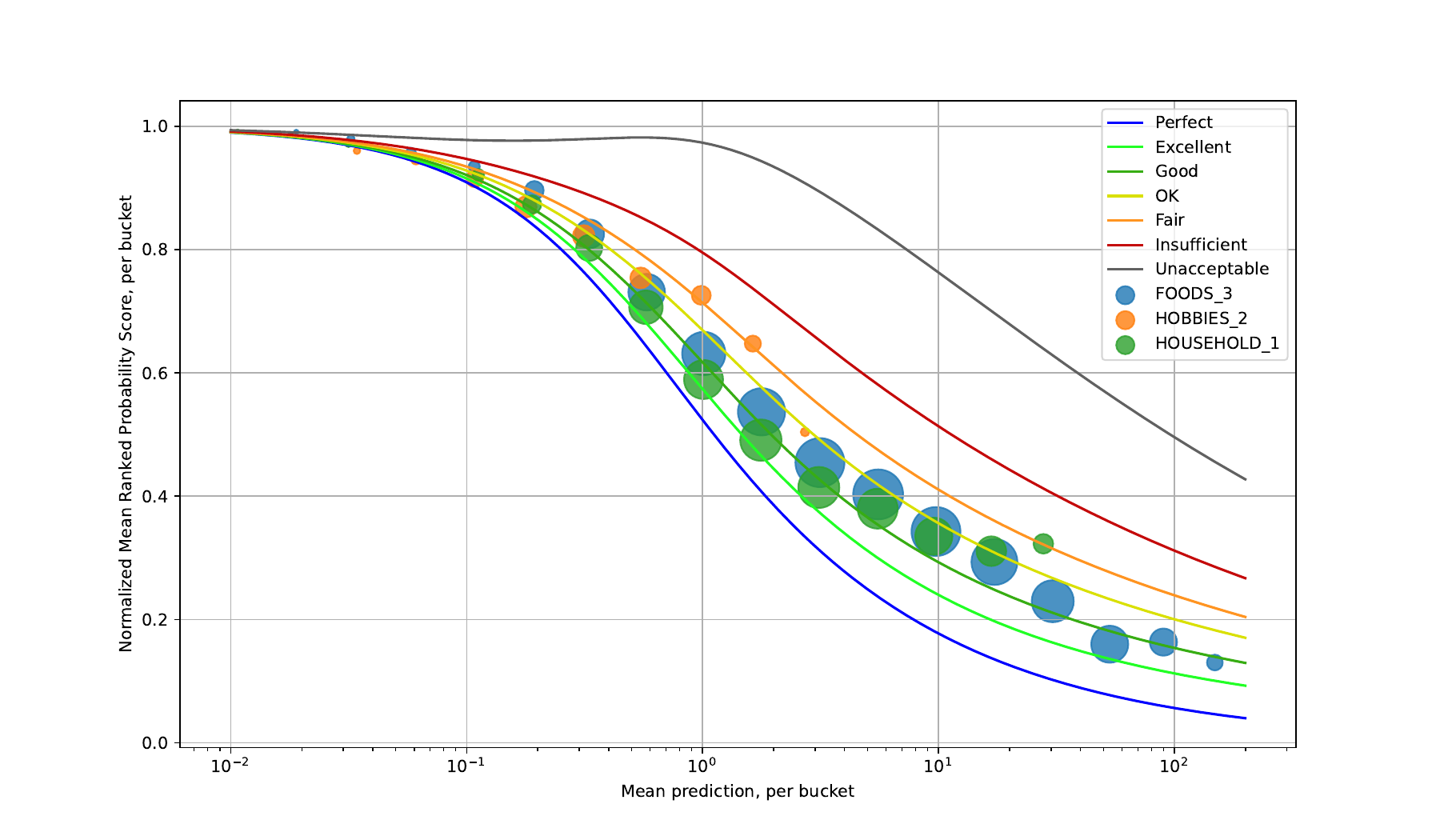}
\caption{ \label{categories_bucket} Prediction-bucketed representation of 
noise (NMRPS), for three exemplary departments (LightGBM model). Circle sizes reflect the square root of the total number of sales contained in the respective buckets. }
\end{figure}

The bucket scatter plot is shown for LightGBM for three selected departments in Figure \ref{categories_bucket}. Table \ref{M5PG} summarizes the achieved NMRPS and overall ratings. 
\begin{table}
{\footnotesize
\begin{tabular}{l|l|l|l|l|l|l}
Department &   NMRPS & $S_{\text{overall}}$(NMRPS)  &   Quality (NMRPS)  & $S_{\text{overall}}$(Bias) & Quality (Bias)  & Bias factor \\ \hline
Foods 3 &  42\%  & 64.9\%  &  Good  &  91.3 \% & Excellent &  1.0 \\
Household 1 &  47\% &  75.6\% & Excellent &  83.7 \% & Excellent & 1.0249\\
Hobbies 2  &   80\%  & 63.5\% &  Good &  46.3\%  & OK & 0.8553 \\
\end{tabular}
}
\caption{\label{M5PG} Scaling-aware ratings on departments level, for the LightGMB model applied to the M5 competition dataset. The ratings summarize the prediction-bucketed representation of Figure \ref{categories_bucket}.}
\end{table}
The comparison of unscaled metrics and scaling-aware ratings per department adds substantial insight: From the NMRPS alone, one cannot make any statement about which department is performing better or worse. Using the rating $S_{\text{overall}}(\text{NMRPS})$, and from the lower panel of Figure \ref{categories_bucket}, we see that ``Household 1'' has less noise than ``Foods 3'', and both these departments perform better than ``Hobbies 2''. ``Foods 3'' enjoys a lack of overall bias (quotient 1.0) and an excellent bias rating (91.3\%), reflecting the lack of substantial bias in every bucket. The largest predictions in ``Household 1'' are clearly deteriorated in both the bias and the noise plots. In ``Foods 3'', only the small predictions ($r<0.3$) stand out as being under-forecasting, a general minor mis-calibration of the LightGBM model (see upper panel of Figure \ref{m5application}).

The systematic negative bias in ``Hobbies 2'' is reflected by the mediocre bias score (46.3\%), while the rating on NMRPS is benevolent (63.5\%) -- a systematic negative bias shifts the reference values for noise to larger values. For the overall KPIs, the reference values of normalized metrics (WMAPE and NMRPS) are rather benevolent, while absolute metrics (MAE and MRPS) are  more strict (the KPI ends up in  the ``unacceptable'' and ``insufficient'' range, respectively). 
When bias is negligible, the ratings of the different departments in Table \ref{M5PG} speak a clear message: ``Household 1'' performs better than ``Foods 3'', while the scaling-unaware MAE, MRPS, WMAPE, NMRPS only provide inconclusive, mixed messages.

\subsection{Demand Edge for Retail predictions at Sainsbury's} 
\label{sainsburys}
The M5 dataset discussion in the previous section has the great advantage to be reproducible. On the other hand, one deals with a curated dataset that does not contain all data that are used in production (e.g.,~the price of the product). To complement our discussion and prove the applicability of our method in practice, we apply the scaling-aware rating to the predictions of the Demand Edge for Retail forecasts produced by Blue Yonder for the UK retailer Sainsbury's. Demand Edge for Retail is based on a causal machine-learning model that computes the contribution of features such as price, weather, day-of-the-week, holiday, promotion and many others to the expected demand of a product \citep{Wick2019,Wick2020}. It produces a forecast that consists of a prediction for the expected mean as well as for the variance, technical details can be found in \citep{cyclicboostinglanding,cyclicboostingexample,cyclicboosting}. To simplify the discussion and make it comparable to the M5 dataset discussion, we only use the predicted mean, and interpret it as predicted mean of the Poisson distribution. 

We analyse the next-day-forecasts (horizon 1) for Monday, October 2nd to Saturday, October 28th, 2023, excluding Sundays. We exclude items that are not sold in units of pieces but volume (liters) or mass (KGs), predictions that are smaller than 0.1 / product-location-day, and unsold product-day-location combinations for which the prediction was 20 or larger, for which we can safely assume an out-of-stock-situation.  
After data cleaning, we remain with more than 250 million product-location-day combinations, corresponding to more than 500 million sold items. Global model over-prediction is less than 1\%. 

\begin{figure}
\centering
\includegraphics[width=0.87\textwidth]{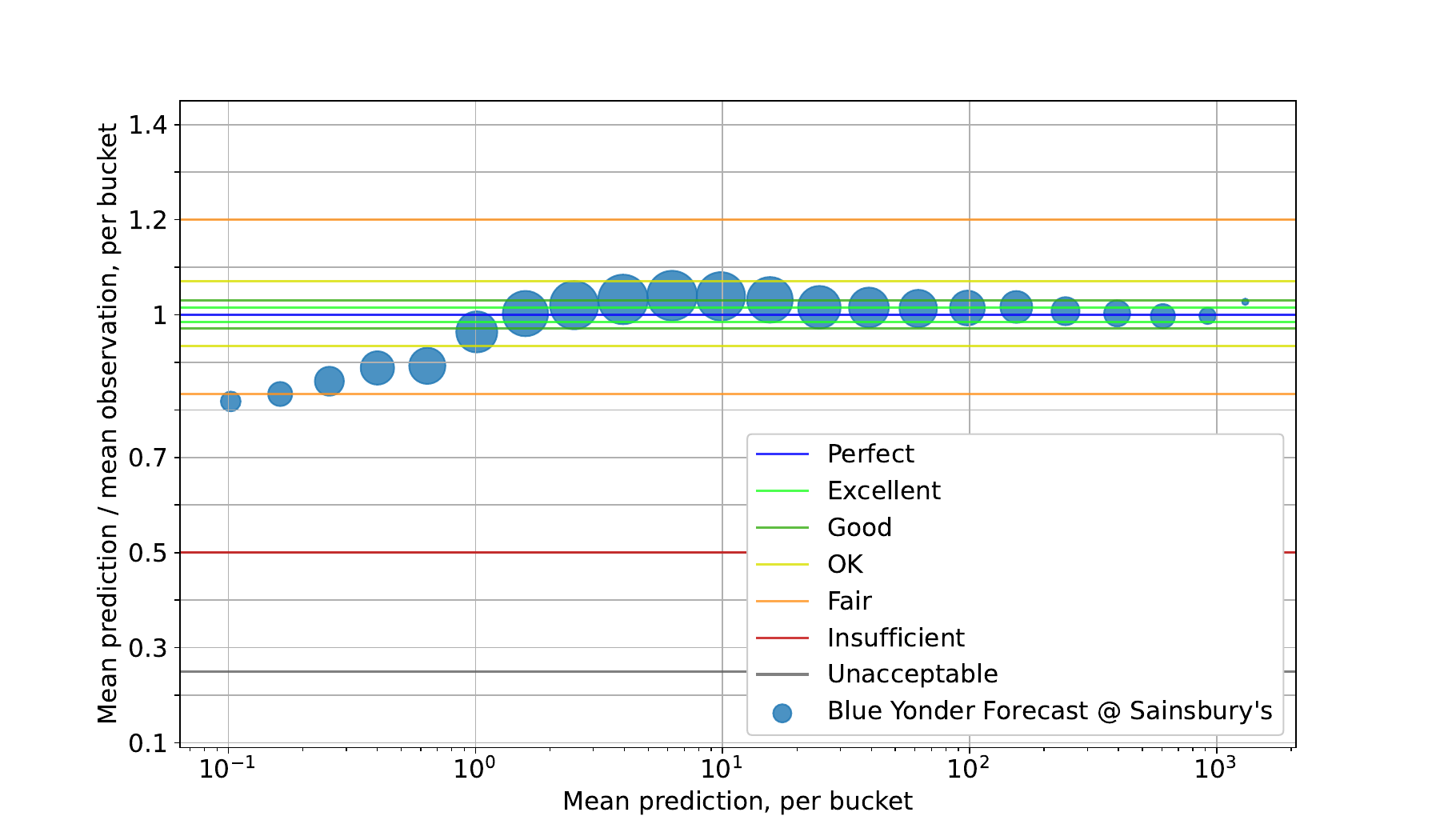}
\includegraphics[width=0.87\textwidth]{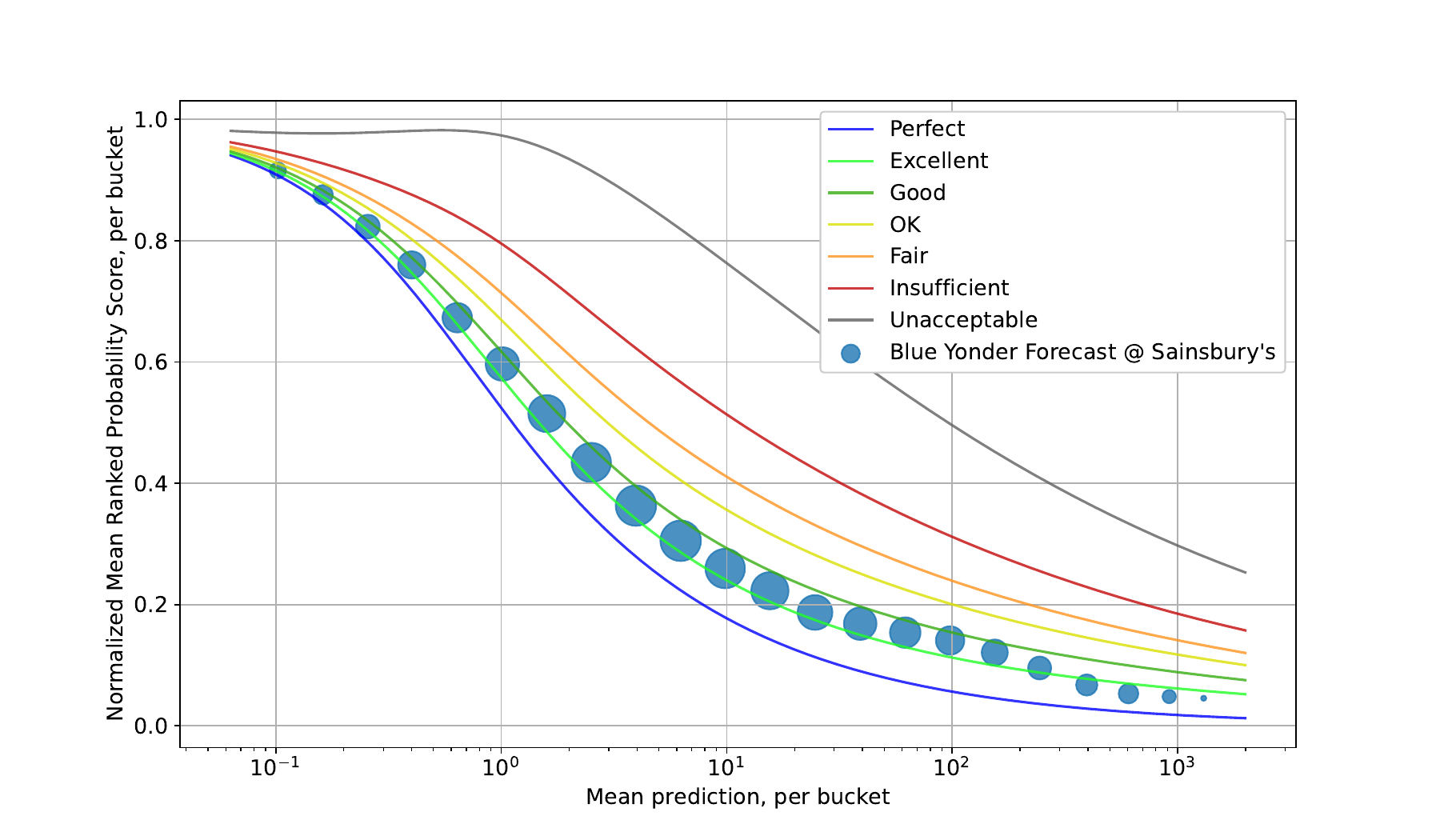}
\caption{ \label{LDE_predictions} Prediction-bucketed representation of bias (upper panel) and noise (NMRPS, lower panel), for Demand Edge for Retail predictions by Blue Yonder for UK retailer Sainsbury's. Circle sizes reflect the square-root of total number of sales contained in the respective buckets. }
\end{figure}

Figure \ref{LDE_predictions} shows the bucketed scatter plots for bias and noise for the Sainsbury's dataset. It confirms the calibration of the forecast (outside the ultra-slow-movers), and the only slightly super-Poissonian noise over three orders of magnitude. Our approach confirms that, on a global level, the prediction is comparable to what could possibly be expected from first principles \citep{Tichy_tech_2023}, as discussed in Section \ref{poisson_sharpest}. This is also reflected by the excellent values $S_{\text{overall}}(\text{NMRPS}) = 84.7 \%$ and  $S_{\text{overall}}(\text{Bias})=78.7 \%$. 

A more in-depth investigation confirms that the reasons for non-Poissonian behavior described in Section \ref{corrections_poisson} above are indeed relevant in this scenario (there are, for example, substantial ``multi-buy'' events when customers buy several pieces of one item). A thorough quantitative understanding of the mild departure from the Poisson ideal case based on those effects would be highly appreciated, but remains beyond the scope of the current work.

\begin{figure}
\centering
\includegraphics[width=0.87\textwidth]{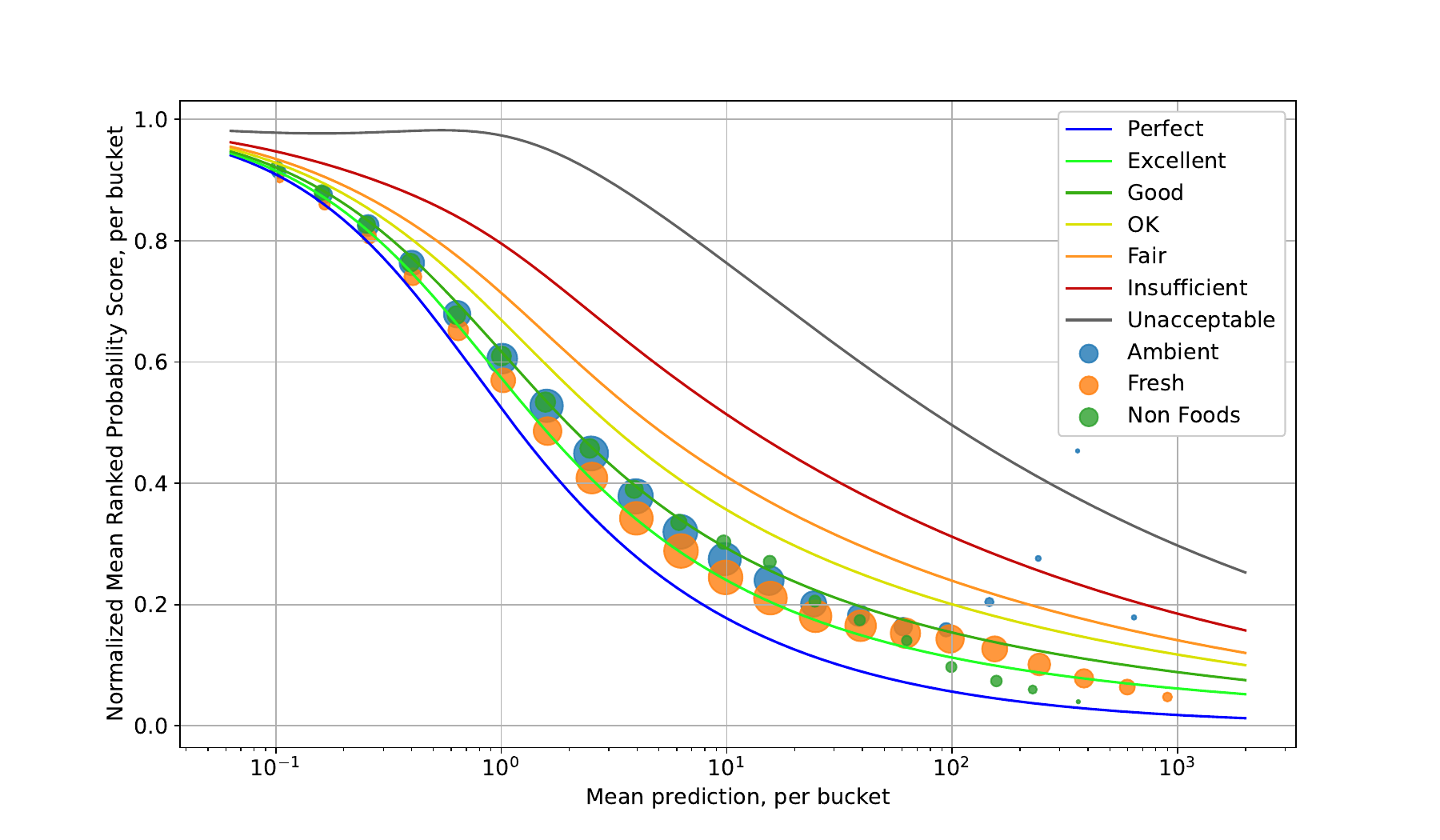}
\caption{ \label{LDE_predictions_pg} Prediction-bucketed representation of  noise (NMRPS), for Demand Edge for Retail predictions by Blue Yonder for UK retailer Sainsbury's, differentiated by the highest-level product group. Circle sizes reflect the square-root of total number of sales contained in the respective buckets.}
\end{figure}

One can segregate data by different dimensions to ask questions as the following: Are certain product groups/stores/weekdays predicted better than others? All of these comparisons risk to induce the na\"ive scaling-trap, which is resolved by our approach. As an example, we differentiate by the highest-level product group and focus on the three largest assortments (Fresh, Ambient, Non foods). The resulting metrics are shown in Table \ref{kpis_sainsbury_pg}. 
\begin{table}
\begin{tabular}{r| r | r r r r r r r}
		                           & All & Fresh  &  Ambient & Non foods \\ \hline
Normalized sales/item &  $100\%$ & $ 200\%$ &  $75\%$ &  $29\%$ \\
Bias factor $\sum_j r_j / \sum_j s_j$  & 1.0092 &  1.0087 & 1.0095 & 1.0100 \\
MAE&   1.108558 &1.724693 & 1.009924 & 0.504811\\
WMAPE&  0.479665 &  0.373344 & 0.58433 &0.751647 \\
MRPS& 0.805966  & 1.252255 &0.732779 &0.374088  \\
NMRPS &0.348735 &  0.271075 &  0.423977 & 0.557006 \\
\end{tabular}
\caption{\label{kpis_sainsbury_pg} Metrics for Blue Yonder's Demand Edge for Retail forecasts at Sainsbury's. Due to non-disclosure reasons, we neither show the absolute total sales nor the sales per item-location-day, but state the sales per product-location-day normalized to the overall mean velocity. Similarly to Table \ref{KPI_benchmark}, the scaling-unaware metrics lead to ambiguous interpretations: Judging from the MAE and MRPS, one would focus to improve the Fresh assortment, based on WMAPE and NMRPS, one would investigate Non foods.}
\end{table}
Together with Figure \ref{LDE_predictions_pg}, we see that the overall numeric metrics (e.g., WMAPE, NMRPS) are misleading if used directly, since Fresh tends to sell at about twice the average velocity -- one more example of the na\"ive scaling trap. Once we apply our bucketed evaluation summarized by Table \ref{sainsburystablepg}, we see that all three groups maintain comparable, near-Poisson performance, earning ``Good'' or ``Excellent'' ratings. We use the same parameters for ratings as for the M5 competition to make the numbers comparable; Blue Yonder uses internally different sets of industry-specific parameters.

\begin{table}
\vspace{0.35cm}
{\small
\begin{tabular}{l|ll|ll}
Product group &  $S_{\text{overall}}$(NMRPS)  &   Quality (NMRPS)  & $S_{\text{overall}}$(Bias) & Quality (Bias)  \\ \hline
Fresh &   88.7\%  &  Excellent  &  84.3\% & Excellent \\
Ambient & 80.2\% & Excellent &   72.5\% &  Good  \\
Non Foods  & 79.3\% &  Excellent &  64.4\%  & Good  \\
\end{tabular}
}
\caption{\label{sainsburystablepg} Scaling-aware ratings for the three main product groups at Sainsbury's, for Blue Yonder's Luminate Demand Edge forecasting solution, summarizing the prediction-bucketed representation of Figure \ref{LDE_predictions_pg}.}
\end{table}

In practice, applying scaling-aware forecast rating at Sainsbury's has resolved many ambiguous situations in which stakeholders were unsure whether certain achieved KPIs were expected or out of range. When introducing model changes, the prediction-bucketed representation was handy to give confidence that changes have a global positive impact, across all velocities. For the forecast provider Blue Yonder, adhering to a clear methodology on how to judge models has streamlined and improved support capabilities. For the forecast consumer Sainsbury's, the scaling-aware analyses have helped tackling data quality problems in a focused manner to improve the forecast, and given the visibility and confidence of how it is performing.

These examples show that applying scaling-aware rating on different data-subsets such as departments, locations, days-of-the week helps focus development efforts on those specific forecasts that exhibit the largest gap to the Poisson ideal case. These are often \emph{not} those that exhibit extremal values of scaling-unaware metrics.

\section{Conclusions and outlook} \label{sec:conclusions}
Applied forecast judgement methodology should reconcile prevailing business practice  \citep{Gartner_FA_target,Bower2023} with the state of the art in probabilistic forecasting \citep{https://doi.org/10.1111/j.1541-0420.2009.01191.x,Gneiting2011,Kolassa2016,Wei2014,GneitingKatzfuss}. 
Scaling-aware forecast rating provides a framework to tackle the problems of benchmarking and precision scaling in a systematic way. By adapting the parametrization introduced in Section \ref{scaling_imperfect_reference}, one can apply  scaling-aware forecast rating to different forecasting problems, e.g.~to industries in which bad data quality is common or certain influencing factors are known to be uncontrollable.

\subsection{Applicability}
Given a forecast, our method permits users to judge the status quo and whether desired improvements in accuracy are realistic at all: If forecast performance is close to the ideal Poisson prediction (or another distribution known to be the sharpest possible for the task at hand), the ask for a substantial improvement needs to be refused. If a gap to the Poissonian case emerges, the reason for this gap needs to be understood. Forecast creators should then elaborate whether it is a realistic expectation to push the forecast closer to the Poisson ideal. 

When several models compete, using scaling-aware methods makes their comparison more transparent and insightful: A given improvement of a metric can be understood holistically by analysing the respective bucket scatter plot as in Figures \ref{m5application}, \ref{categories_bucket}, \ref{LDE_predictions}, \ref{LDE_predictions_pg}. Two competing models can be placed onto a meaningful scale from ``perfect'' to ``unacceptable'', providing context and orientation whether a certain improvement is substantial and solid, or little more than a mere coincidence.

The greatest added value of our method lies in dataset comparison: A comparison of the values of scaling-unaware metrics such as MAE, WMAPE or (N)MRPS  for two sub-datasets of the same model is useless, since the na\"ive scaling trap  almost always snaps. Comparing scaling-aware ratings for different departments or locations instead allows users to focus development efforts on those data subsets for which the gap to the ideal behavior is most pronounced. For example, when monitoring forecasting performance in time, usual metrics will change due to increasing or decreasing overall volume. Thresholds on scaling-unaware metrics (``if WMAPE increases above 70\%, trigger an alarm!'') are prone to false positives and false negatives, which scaling-aware ratings remedy.

\subsection{Future venues}
Our proposed method -- computing expectation values of metrics, and bucketing by similar predictions -- can be applied to all probabilistic forecasts for which one can characterize the ideal case. Within the application in retail, several generalizations are thinkable: In a business context, it is advisable to judge the forecast by its actual economic impact \citep{SYNTETOS2010134}, which can be rated in the context of what could be possibly achieved. A certain out-of-stock or waste loss, expressed in currency, can then be given a rating (``perfect'', ``excellent'', etc.). The question naturally emerges to which extent human overrides of forecasts \citep{Khosrowabadi2022} intuitively incorporate precision scaling. 

The exponent $\gamma=1.5$ in the parametrization in Eq.~(\ref{variance_general_scaling}) works remarkably well and quite universally, for many different applications and industries (compare the M5-analysis to the Sainsbury's analysis -- models on both datasets follow the shape of the reference lines across several orders of magnitude). Nevertheless, a better theoretical explanation of that value would be appreciated. Our hope is fuelled by the fact that scaling laws often reflect underlying system structures  \citep{scalewest,ATHANASOPOULOS2022}.

In the case of finite capacity, which can result in stock-outs, one can bucket the data both by the predicted uncensored selling rate and by the predicted probability to go into stock-out \cite{tichy_2024_censoring}. This then allows again to compare the expected value of the metric to the achieved one.

A forecast judgement summarizes a vast number of predictions and observations into a few summary statistics. Without having set a rigorous benchmark, achieved metrics lack context and evade interpretability. Scaling-aware forecast rating, applied on a bias- and on a noise-related metric, reduces the likelihood of such paradoxical situations by setting a context and avoiding the \emph{na\"ive scaling trap}. We hope that it will help users handle and judge forecasts in practice, in a statistically corroborated way.

\section*{Acknowledgements}
The authors would like to acknowledge helpful discussions with Ulla Gebbert, S\"o{}nke Niekamp, Hannes Sieling and Martin Wong, and comments on the manuscript by Martin G\"o{}rner, Alexander Berkner and Pascal Jordan. They are grateful to Sainsbury's to allow us to showcase our method on their productive data. 

\section*{Declaration of competing interest}
Blue Yonder has filed patent applications covering the intellectual property associated with this paper.

\section*{Data Availability}
The data of the M5 competition that support the findings of this study are openly available at \cite{kaggle}. Blue Yonder and Sainsbury's did not give written consent for their data to be shared publicly.

\appendix

\section{Heuristic algorithm for non-parametric Expectation Maximization} \label{heuristic_algorithm}
To discuss our rating methodology on ``perfect'' Poissonian forecasts for a given dataset of sales values $s_j$, we construct baseline predictions using the following heuristic, which is equivalent to Expectation Maximization: Given the observed sales $s_j$, we assume that these integer outcomes are the results of Poisson processes with rates $t_j$, i.e.,~that Eq.~(3) of the main text holds. Given the known observations $s_j$, we need to evaluate $P(t | s)$, i.e.,~the conditional probability to have predicted $t$ given that we do observe $s$, to create plausible predictions $t_j$. Remember that Eq.~(3) of the main text states the conditional probability to observe $s$, given $t$. Bayes' rule gives
\begin{equation}
P(t | s) = \frac{P_{\text{Poi}}(s | t) P_{\text{rate}}(t) }{P_{\text{observation}}(s)} .
\end{equation}
 The probability of a certain outcome $P_{\text{observation}}(s)$ is the empirical frequency of that value $s$ in the dataset. To evaluate $P(t|s)$, we need an approximation to $P_{\text{rate}}(t)$, the prior probability density distribution of the rates. 

We approximate this prior by starting with a an exponential distribution,
\begin{equation}
P_0(t) = \frac{1}{\langle s \rangle} e^{-\frac{t }{ \langle s \rangle} } ,
\end{equation}
whose mean value is set to match the mean observed sales 
\begin{equation} 
\langle s \rangle  =  \sum_{s=0}^{\infty} s P_{\text{observation}}(s) .
\end{equation}
We denote by $P_{\text{observation}}^{(k)}(s)$ the resulting probability to observe $s$, given the prior for the rates $P_k(t)$, i.e. 
\begin{equation}
P_{\text{observation}}^{(k)}(s) = \int_{t=0}^{\infty} P_k(t) P_{\text{Poi}}(s | t) ,
\end{equation}
where the index $k$ will denote the iterations of the algorithm. For the first guess and before the first iteration,  $P_{\text{observation}}^{(0)}(s)$ and $P_{\text{observation}}(s)$ typically differ  substantially. 

We apply the following update rule to improve the probability density function $P_k(t)$ iteratively:
\begin{equation}
 \tilde P_{k+1}(t) = P_k(t) \sum_{s=0}^{\infty}  P_{\text{Poi}}(s|t) \frac{P_{\text{observation}}(s)}{P^{(k)}_{\text{observation}}(s)}, \label{updaterule} 
\end{equation}
where, intuitively speaking, we boost the probability density that contributes to those observations that are currently under-predicted. The updated probability distribution $ \tilde P_{k+1}(t)$ is unnormalized, such that we apply 
\begin{equation}
P_{k+1}(t) =  \frac{\tilde P_{k+1}(t)}{\int_{x=0}^{\infty} \text{d}x \tilde P_{k+1}(x) } ,
\end{equation}
before the update rule (\ref{updaterule}) is applied again. 

To provide a set of plausible predictions $t_j$, we sample one prediction $t_j$ for each observation $s_j$ via  $t_j \sim P(t | s_j)$. Applying the iterative procedure on all data categories (product groups, locations...) separately, we obtain unbiased Poissonian forecasts that could have yielded the set of sales $s_j$. We have experienced that a dozen iterations is typically sufficient to reach a good artificial forecast, i.e.,~a forecast for which $P^{(k)}_{\text{observation}}(s) \approx P_{\text{observation}}(s)$ while, for each prediction $t$, the resulting sales $s$ are Poisson-distributed (see Figure \ref{fig:fig_histograms}). In practice, we discretize the space of rates $t$ into a granular array of length $G=5'000$, $t_0 \dots t_{G}$, such that (\ref{updaterule}) is performed on all $P_{k}(t_0) \dots P_{k}(t_G)$ in a numerical fashion.

\subsection{Treatment of improbable observations}
In the update rule (\ref{updaterule}), the cases $P^{(k)}_{\text{observation}}(s)=0$ and/or $P_{\text{observation}}(s)=0$ need special attention.
\begin{itemize}
\item When $P_{\text{observation}}^{(k)}(s)=P_{\text{observation}}(s)=0$, we set the quotient of ${P_{\text{observation}}(0)}$ and ${P^{(k)}_{\text{observation}}(0)}$  to 1: The probability 0 for the observation $s$ predicted by $P_{\text{observation}}^{(k)}(s)$ matches the empirical  frequency $P_{\text{observation}}(s)$, no adjustment is needed. 
\item When $P_{\text{observation}}(s)=0$ but $P_{\text{observation}}^{(k)}(s)>0$, the quotient is kept at 0: The empirical frequency is 0, hence, the components that contribute to that observation can be decreased. 
\item When $P_{\text{observation}}(s)>0$ but $P_{\text{observation}}^{(k)}(s)=0$ (or very close to 0), we need to ``boost'' components that contribute to that observation $s$, hence we set the otherwise undefined (or very large) quotient to 2. 
\end{itemize}

\subsection{Application to M5 dataset}
\begin{figure}
\centering
\includegraphics[width=\textwidth]{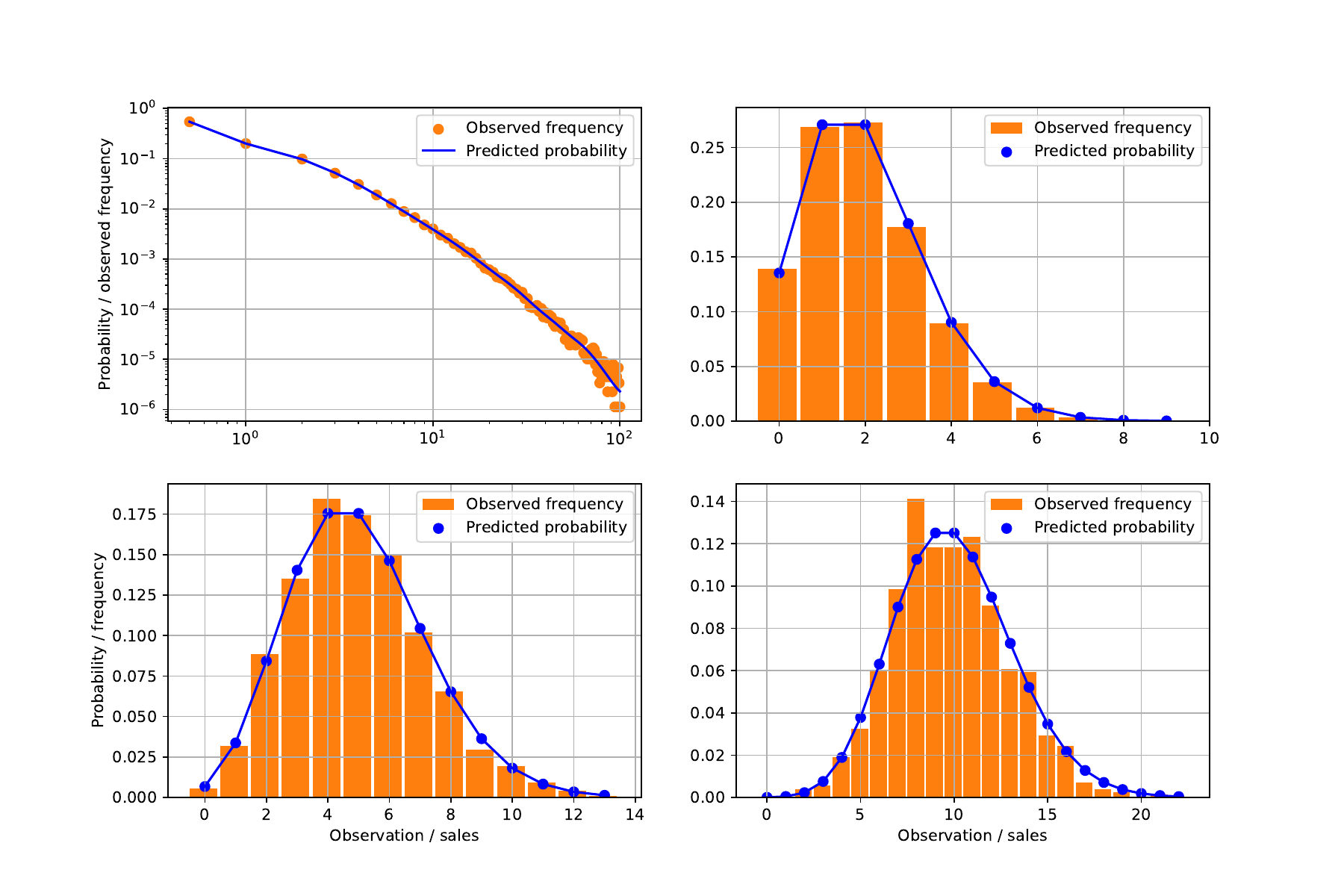}
\caption{ \label{fig:fig_histograms} Calibration diagrams for ideal baseline model, based on the heuristic of \ref{heuristic_algorithm}: Distribution of observed sales frequency (orange) vs. predicted probability (blue, this color refers to the ideal Poisson case throughout this article). Upper left: Overall marginal calibration diagram (for visual convenience, we mapped the observation 0 to 0.5 to fit into the log-log-scale). Upper right: Cut on $1.8<r<2.2$. Lower left: Cut on $4.8<r<5.2$. Lower right: Cut on $9.8<r<10.2$.}
\end{figure}

The calibration diagrams of Figure \ref{fig:fig_histograms} confirm that the baseline $\vec r^{\text{baseline}}$ that we constructed by following the EM procedure is an excellent baseline model, since the unconditioned and the conditioned distributions of actually observed sales $s$ match the Poisson prediction well. 

\FloatBarrier
\section{Summarizing buckets to overall ratings} \label{appendix_summarizing}
We elaborate on the summary of bucket-wise to overall ratings that we sketched in Section 3.4 of the main text.  We assume to hold a set of prediction-stratified buckets indexed by $R$, the rounded logarithm of the rate defined. Each bucket contains $n_R$ prediction/observation pairs, denoted by $\vec r^{(R)}, \vec s^{(R)}$. For each bucket, we compute  the total and the average prediction
\begin{eqnarray}
r_{\text{total}}^{(R)}&=&\sum_{j, R_j=R} r_j, \\
r_{\text{mean}}^{(R)} &=&r_{\text{total}}^{(R)} / n_R,
\end{eqnarray}
 the total and the average observation
\begin{eqnarray}
s_{\text{total}}^{(R)}&=&\sum_{j, R_j=R} s_j, \\
s_{\text{mean}}^{(R)} &=& s_{\text{total}}^{(R)}/n_R
\end{eqnarray}
 and the achieved metrics $M^{(R)}_{\text{actual}}=M(\vec r^{(R)}, \vec s^{(R)} )$. By plotting the achieved metrics against the predicted rates in the context of the metrics reference values (i.e.,~using Figure 5 of the main text as a background context), one can visually extract useful insights on model quality and possible improvements, as exemplified in Section 4 of the main text.

In certain situations, it might be desirable to condense and summarize the information contained in a set of buckets by a single number. There are two possibilities to perform this summary: aggregate the ratings of the metric per bucket to yield an overall score, or judge the overall aggregated metric, e.g.~the overall achieved WMAPE.

\subsection{Aggregate bucket ratings} \label{aggregate_bucket_ratings} Given an evaluation metric $M$, we compute the ratings of the individual buckets and average them to an overall rating: For each bucket $R$, the achieved metric value $M^{(R)}_{\text{actual}}$ is compared to  the reference values of the metric for the different ratings, i.e.,~to
\begin{equation}
M^{(R)}_{\text{perfect}},~M^{(R)}_{\text{excellent}},\dots , M^{(R)}_{\text{unacceptable}}. \label{refvaluesperbucket}
\end{equation}
 Let the metric $M$ be a loss function, such that larger values are worse, and $M^{(R)}_{\text{actual}} > M^{(R)}_{\text{unacceptable}}$ reveals some dramatic problem of the forecast in the respective bucket. 

To define numerical scores that allow an average of the ratings resulting in some overall score $S_{\text{overall}}$, we map quality categories (``perfect'', ``excellent'', \dots ,``unacceptable'') to percentage ratings $S_{\text{quality}}$, using a scale between 100\% (best) to 0\% (worst):

\vspace{0.35cm}
{\small
\begin{tabular}{r|lllllll}
Quality & Perfect & Excellent & Good & OK & Fair & Insufficient & Unacceptable \\ \hline
 $S_{\text{quality}}$ & $>91.67 \%$ & $>75\%$ & $>58.33 \%$ & $>41.67\%$ & $>25\%$ & $>0.083\%$   &$ <0.083 \% $
\end{tabular}
}
\vspace{0.35cm}

For each bucket, the achieved metric $M^{(R)}_{\text{actual}}$ implies some $M^{(R)}_{\text{upper}} <M^{(R)}_{\text{actual}} < M^{(R)}_{\text{lower}}$  defining the two neighboring qualities, with percentage ratings $S_{\text{upper}} > S_{\text{lower}}$. For the bias, we map negative bias values $ r_{\text{total}}^{(R)} / s_{\text{total}}^{(R)}=b<1$ to $1/b > 1$ to achieve the monotonicity of our rating (for example, we rate a bias of 1.1 as being equivalent to a bias of $1/1.1\approx 0.91$).

We set the percentage rating of the bucket to the linear interpolation of the ratings achieved by the neighboring buckets:
\begin{equation}
S^{(R)} = S_{\text{lower}} + (S_{\text{upper}} -S_{\text{lower}} ) \frac{M^{(R)}_{\text{actual}} - M^{(R)}_{\text{upper}}}{M^{(R)}_{\text{lower}} - M^{(R)}_{\text{upper}}} , \label{score_per_bucket}
\end{equation}
where this pragmatic choice might be replaced by splines or other interpolation methods.
Eq.~(\ref{score_per_bucket}) assigns a numerical score to each bucket. A score of 100\% is achieved when the metric computed in the bucket matches exactly the value expected under the Poisson distribution. A score of 0.083\% is awarded when the metric matches the value $M^{(R)}_{\text{unacceptable}}$ that is achieved by a distribution affected by ``unacceptable'' noise (according to the rating). The worst score, 0\%, is reached when twice the ``unacceptable'' value is attained for loss functions. The numerical score thus quantifies ``how far from the ideal'' the bucket lies in the prediction-metric-area, and how that shall be interpreted operationally.

The overall rating $S_{\text{overall}} (\vec r, \vec s)$ is defined by the weighted mean of the $S^{(R)}$ that are awarded to the buckets:
\begin{equation}
S_{\text{overall}}(\vec r, \vec s) = \frac{ \sum_{R} S^{(R)} \cdot \text{max}(s^{(R)}_{\text{total}} , r^{(R)}_{\text{total}} ) }{ \sum_{R} \text{max}(s^{(R)}_{\text{total}} , r^{(R)}_{\text{total}} )} \label{overall_rating}
\end{equation}
where each bucket is weighted by either its total prediction or its total observation, depending on which is larger. We do not merely weight by total bucket sales or by total bucket predictions, but by bucket-wise maximum to avoid neglecting buckets with extreme over- or under-forecasting. The resulting $S_{\text{overall}}$ is then the scaling-aware rating of the overall set of predictions and observations, which can corroborate operational decisions. 

\subsection{Rate the overall aggregated metric} \label{subsubsec:rate_overall}
The overall metric $M(\vec r, \vec s)$ is embedded into a context by computing reference values $M_{\text{perfect}}(\vec r)$,  $M_{\text{excellent}}(\vec r)$, \dots $M_{\text{unacceptable}}(\vec r)$ as follows: For a given set of buckets, we compute the overall metric $M_{\text{quality}}(\vec r)$ which would have been achieved if all buckets $R$ had achieved the $M^{(R)}_{\text{quality}}$ from Eq.~(\ref{refvaluesperbucket}). 
This allows the user to corroborate statements such as ``an overall WMAPE of 44\% was achieved, which is between good (38\%) and OK (46\%)''.

The rating of the overall aggregated metric and the weighted mean rating of the buckets will often yield similar conclusions. They are, however, answering a different question and can, in certain cases, deviate substantially: Consider a forecast that is unbiased overall,
\begin{equation}
\sum_R r^{(R)}_{\text{total}} \approx \sum_R s^{(R)}_{\text{total}} .
\end{equation}
The overall aggregated bias is, thus, excellent. However, it might still be the case that slow-movers are heavily underforecasted while fast-movers are overforecasted (or vice versa), such that
\begin{equation}
\forall R: r^{(R)}_{\text{total}} \gg  s^{(R)}_{\text{total}} \text{ or }r^{(R)}_{\text{total}} \ll  s^{(R)}_{\text{total}}
\end{equation}
The scaling-aware rating then yields a critical value (such as ``fair'', ``insufficient''...) because each bucket is biased, while the overall aggregated bias seems to be excellent.

%\section*{References}

%\bibliographystyle{rss}
%\bibliography{library.bib}

\end{document}